\documentclass[english,12pt,prd,showpacs,nofootinbib]{revtex4}
\usepackage{bm}
\usepackage{graphics}
\usepackage{rotating}
\usepackage{epsfig}

\usepackage[T1]{fontenc}
\usepackage[latin9]{inputenc}
\usepackage{array}
\usepackage{longtable}
\usepackage{graphicx}
\usepackage{amssymb}
\usepackage{babel}

\def\bma{\left( \begin{array} }
\def\ema{\end{array} \right)}
\def\gsim{~{\rlap{\lower 3.5pt\hbox{$\mathchar\sim$}}\raise 1pt\hbox{$>$}}\,}
\def\lsim{~{\rlap{\lower 3.5pt\hbox{$\mathchar\sim$}}\raise 1pt\hbox{$<$}}\,}

\newcommand{\MNS}{{\text{MNS}}}

\def\bmaT{\left(\begin{array}{ccc}}
\def\emaT{\end{array}\right)}
\let\jnfont=\rm
\def\NPB{\jnfont Nucl.\ Phys.\ B}
\def\PLB{\jnfont Phys.\ Lett.\ B}
\def\EPJC{\jnfont Euro.\ Phys.\ J.\ C}
\def\PRD{\jnfont Phys.\ Rev.\ D}
\def\PRL {\jnfont Phys.\ Rev.\ Lett.}

\def\JHEP{\jnfont J. High \ Ener.\  Phys.}
\def\RMP{\jnfont  Rev. Mod. Phys.}

\def\PLB{\jnfont Phys. Lett. B}

\def\Rv{\not{\hbox{\kern-1pt $R$}}}
\def\p{\not{\hbox{\kern-3pt $p$}}}

\begin{document}
\begin{flushright}{ arXiv:1309.5495}\end{flushright}
\title{The lepton flavor violating signal of the charged scalar $\phi^\pm$ and $\phi^{\pm\pm}$ in
photon-photon collision at the ILC }  
\author{Guo-Li Liu$^{1,3}$, Fei Wang$^{1,3}$, Qing-Guo Zeng$^{2,4}$}
\affiliation{ $^1$ Physics Department, Zhengzhou University, Henan,
450001, China \\
 $^2$Department of  Physics, Shangqiu Normal University, Shangqiu  476000,  China \\
$^3$ Kavli Institute for Theoretical Physics, Academia Sinica,
Beijing 100190, China \\
 $^4$Department of Physics, Liaoning Normal University, Dalian 116029,  China }
\vspace{15mm}
\begin{abstract}
The hitherto unconstrained lepton flavor mixing, induced by the new
charged scalar $\phi^\pm$ and $\phi^{\pm\pm}$ predicted
by many new physics models such as Higgs triplet models, may
lead to the lepton flavor violating
productions of $\tau\bar \mu$, $\tau\bar e$ and $\mu\bar e$ in
photon-photon collision at the proposed international linear
collider (ILC). 

In this paper, we consider the contributions
of the $\phi^\pm$ and $\phi^{\pm\pm}$ in the context of the Higgs triplet models to the
processes $\gamma\gamma\to l_i\bar l_j$ ($i,~j= e,~\mu,~\tau,~i\neq j$)
and find that they can be good channels
to probe these new physics models.  The lepton flavor violating processes
$\gamma\gamma\to l_i\bar l_j$ ($i,~j= e,~\mu,~\tau,~i\neq j$)
 occur at a high rate due to the large mixing
angle and the large flavor changing coupling,
so, in view of the low standard model backgrounds, they may reach the
detectable level of the ILC for a large part of the parameter space.
 Since the rates predicted by the standard model are far below the
detectable level, these processes may serve as a sensitive probe for
such new physics models.
\begin{flushleft}
\hspace{12mm}\textbf{Keywords:}  new
charged scalar $\phi^\pm$ and $\phi^{\pm\pm}$, lepton flavor
violating processes, photon-photon collision, Higgs triplet model
\end{flushleft}
\end{abstract}


\maketitle

\newpage
\section{\bf Introduction}

Lepton flavor violating (LFV)
interactions are missing in the Standard Model (SM), 
so any observation of the LFV processes would serve as a robust evidence for new physics beyond
the SM. 

Many kinds of models beyond the SM, such as the Higgs triplet models
(HTM)~\cite{Schechter:1980gr,Cheng:1980qt}
predict the presence of charged scalars $\phi^\pm$ and $\phi^{\pm\pm}$.
 Such triplet Higgs fields can induce LFV
processes at the proposed international linear collider (ILC)\cite{ilc-project},
 such as the productions of $\tau\bar
\mu$, $\tau \bar e$ and $\mu \bar e$ via $e^+ e^-$,  $e^- \gamma$
and $\gamma \gamma$ collisions.
 It is noticeable that the productions of  $\tau\bar \mu$, $\tau \bar e$ and
$\mu \bar e$ in $\gamma \gamma$ collision have not been studied in
this scenario. It is also noticeable that
all these LFV processes at the ILC involve the same part of the
parameter space of such new physics models.
 Therefore, it is necessary to compare all these processes
to find out which process is the best to probe these models.

Due to its rather clean environment, the ILC will be an ideal machine to probe new physics.
The LIC is a proposed future $e^+e^-$ collider, designed to
fill $e^+e^-$ collisions at energies from 0.5 to 1 TeV, with the possibility to update to
 3 TeV, which is actually designed to be compact linear
 collider(CILC)\cite{clic-project}.

 In addition to $e^+ e^-$ collision, we can also
realize $\gamma \gamma$ collision\cite{rr-project} in such a collider with the photon beams generated by
the backward Compton scattering of incident electron- and
laser-beams.

 The LFV productions in $\gamma \gamma$ collision may be
more important than those in $e^+ e^-$ collision. 
Firstly, $ e^+e^- \to \ell_i \bar \ell_j$ can be generated by means of the photon s-channel
like $e^+e^- \to  \gamma^* \to \ell_i \bar \ell_j$, with $ S^{\pm\pm}$ and/or $H\pm$ running inside the loop,
which is  at the same order as cross section of the $ \gamma\gamma \to \ell_i \bar \ell_j$.
However, the $e^+e^-$ production is expected to be sub-dominant with respect to
the production from $\gamma\gamma$ collision  since the latter gets the usual logarithmic enhancements induced
by the phase space integration of the $u$- and $t$-channels.
More importantly, compared with the collision in the $e^+e^-$, the lepton
flavor violating productions at the $\gamma\gamma$ collision are
essentially free of any SM irreducible background.
So the LFV productions in the $\gamma\gamma$ collision are a good probe for new physics models.

In this work, we will study the LFV processes $\gamma\gamma \to
\ell_i\bar \ell_j$ ( $\ell_i = e,~\mu~\tau $ and $i\neq j$) which is induced
by the the new charged scalars $\phi^\pm$ and $\phi^{\pm\pm}$ in HTM models.
We calculate the production rates to figure out if they can reach the sensitivity
of the photon-photon collision of the ILC within the allowed parameter space of this scenario.

The work is organized as follows. We will briefly discuss the HTM
models in Section II and III, giving the involved new couplings and the parameters in our calculation.
In Section IV we give the
calculation results in the HTM models and compare them
with other models, such as the supersymmetry, the little Higgs and technicolor models.
Section V is our conclusion.
\section{the Higgs triplet models and the relevant couplings}

Many new physics scenarios predict new particles which lead to significant LFV signals.
For example, the charged scalars $\phi^\pm$ and $\phi^{\pm\pm}$, which are
predicted by various specific models beyond the SM, can lead
to the large tree-level lepton flavor
changing couplings. Such couplings can have significant
contributions to the realization of some LFV processes.  

In the Higgs triplet model~(HTM) \cite{Schechter:1980gr,Cheng:1980qt}
 an extra $SU(2)_L$ isospin scalar triplet is added to the SM state spectrum.
The neutrinos can directly obtain a majorana mass from the triplet through the
gauge invariant Yukawa interaction, in the absence of the right-handed neutrinos.
The Yukawa interactions can be written as \cite{09043640}:
\begin{equation}
{\cal L}=h_{ij}\psi_{iL}^TCi\tau_2\Delta\psi_{jL}+h.c~,
\label{trip_yuk}
\end{equation}
where the coupling $h_{ij} (i,j=1,2,3)$ is complex and symmetric.
$C$ and $\tau_2$ denote the Dirac charge conjugation operator and the second Pauli matrix, respectively.
$\psi_{iL}=(\nu_i, l_i)_L^T$ is the left-handed lepton doublet.
$\Delta$ is a new complex triplet fields of $Y=2$ with a $2\times 2$ representation:
\begin{equation}
\Delta
=\bma{cc}
\Delta^+/\sqrt{2}  & \Delta^{++} \\
\Delta^0       & -\Delta^+/\sqrt{2}
\ema
\end{equation}

The non-zero vacuum expectation value (VEV)  $\langle\Delta^0\rangle$
 of the triplet fields $\Delta$, results in the following neutrino mass matrix: 
\begin{equation}
m_{ij}=2h_{ij}\langle\Delta^0\rangle = \sqrt{2}h_{ij}v_{\Delta}
\label{nu_mass}
\end{equation}

The necessary non-zero $v_{\Delta}$ is generated by the minimization of
the most general $SU(2)\otimes U(1)_Y$ invariant Higgs potential,
which can be written as follows~\cite{Ma:2000wp, 0304069}
(with $\Phi=(\phi^+,\phi^0)^T$):
\begin{eqnarray}
V&=&m^2(\Phi^\dagger\Phi)+\lambda_1(\Phi^\dagger\Phi)^2+M^2
{\rm Tr}(\Delta^\dagger\Delta) +
\lambda_2[{\rm Tr}(\Delta^\dagger\Delta)]^2+ \lambda_3{\rm Det}
(\Delta^\dagger\Delta)  \nonumber \\
&&+\lambda_4(\Phi^\dagger\Phi){\rm Tr}(\Delta^\dagger\Delta)
+\lambda_5(\Phi^\dagger\tau_i\Phi){\rm Tr}(\Delta^\dagger\tau_i
\Delta)+\left(
{1\over \sqrt 2}\mu(\Phi^Ti\tau_2\Delta^\dagger\Phi) + h.c \right)
\label{Higgs_potential}
\end{eqnarray}
For small $v_\Delta/v$,
the expression for $v_\Delta$
resulting from the minimization of $V$ is:
\begin{equation}
v_\Delta \simeq \frac{\mu v^2}{2M^2+(\lambda_4+\lambda_5)v^2} \ .
\label{tripletvev}
\end{equation}

The possibility of the observations of various lepton flavor violating processes
induced by the 
triplet Higgs bosons 
can provide a probe for the
neutrino masses and mixing through the relation (\ref{nu_mass}),
and thus a direct test of the model.

In the HTM models, there are seven Higgs bosons $(H^{++},H^{--},H^+,H^-,H^0,A^0,h^0)$ \cite{09043640}.
The doubly charged $H^{\pm\pm}$ can be identified with a component of the triplet scalar field $\Delta^{\pm\pm}$.
The remaining eigenstates $H^\pm,~ H^0,~A^0,~h^0$  are the mixtures of the
triplet and doublet fields and such mixing is proportional to the triplet VEV and thus small.
The triplet fields are the main component of $H^\pm,~H^0,~A^0$
while $h^0$ is predominantly made up by the
doublet field and act as the SM Higgs boson.
For triplet Higgs bosons masses $M < 1$~TeV, the couplings $h_{ij}$ are
constrained to be $h_{ij}\leq 1 $ or even much smaller than
$1$ by the lepton flavor violating processes
such as $\mu\to e\gamma $, $\tau\to e(\mu)\gamma $,
$\mu\to eee$, and $\tau\to lll$ etc\cite{9511297,0304254,0304069,09043640}.

\section{The Processes and the Parameters Involved}
\begin{figure}[tb]
\begin{center}
\epsfig{file=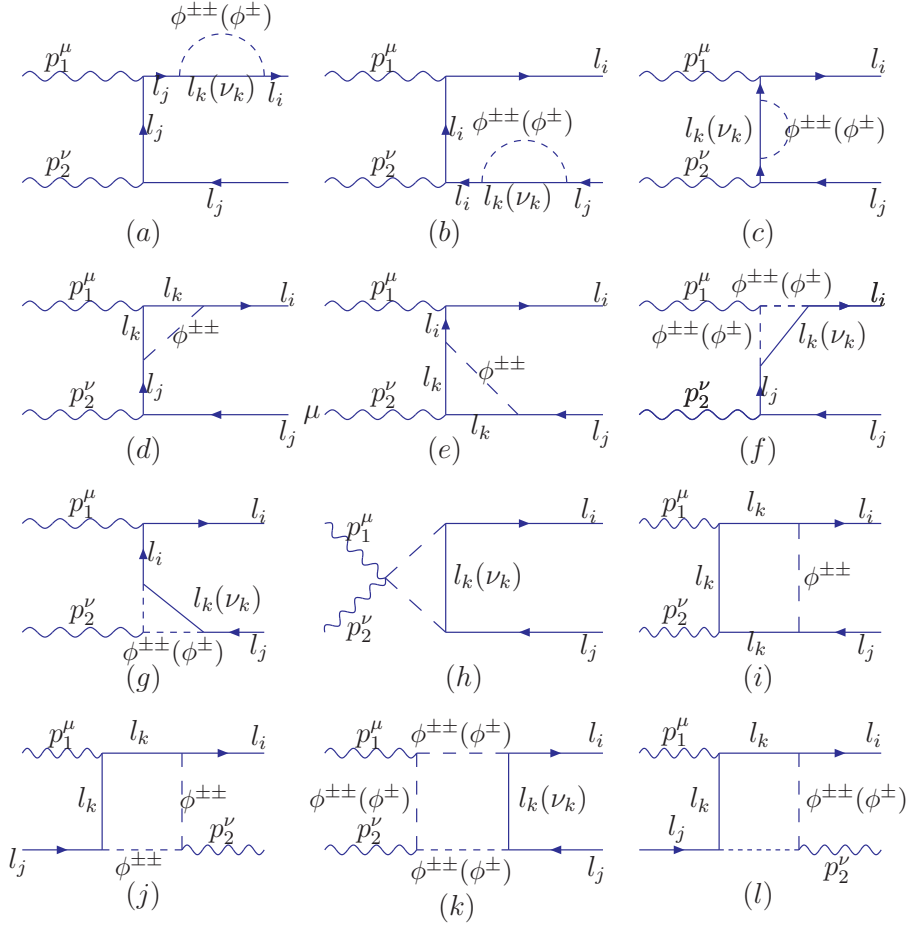,width=12cm} \vspace{-0.7cm}\caption{ Feynman
diagrams contributing to the process $\gamma\gamma \to
\ell_i\ell_j$.} \label{fig1}
\end{center}\vspace{-0.7cm}
\end{figure}

 The Feynman diagrams of the LFV processes $ \gamma\gamma \to
\ell_i\ell_j$ ($i\neq j$ and $\ell_i = e,~\mu,~\tau$) induced by the
charged scalars $\phi^{\pm\pm}$ and $\phi^{\pm}$ are shown in
 Figure~\ref{fig1}. There are s-, t- and u- channel contributions in total
 with the u-channel not shown in Figure~\ref{fig1}.

The gauge invariant amplitude of $\gamma\gamma \to \tau\bar\mu(\bar e)$
induced by the scalars is given by
\begin{eqnarray}
{\cal M}=\frac{1}{2} ~\bar{u}_\tau\Gamma^{\mu\nu}P_Lv_\mu
~\epsilon_\mu(\lambda_1)\epsilon_\nu(\lambda_2)~.
\end{eqnarray}
Explicit form of the tensor $\Gamma^{\mu\nu}$ is determined by detailed
couplings which we do not shown here but has been checked
many times in the program. These amplitudes
contain the Passarino-Veltman one-loop
functions, which are calculated by LoopTools \cite{Hahn}.

For $\gamma \gamma$ collision, where the
photon beams are generated by the backward Compton scattering of
incident electron- and laser-beams just before the interaction
point, through convoluting the cross
section  of $\gamma\gamma$ with the photon beam luminosity
distribution, the events number can be written as:
\begin{eqnarray}\label{n_rr}
N_{\gamma \gamma \to \ell_i \bar\ell_j}&=&\int d\sqrt{s_{\gamma\gamma}}
  \frac{d\cal L_{\gamma\gamma}}{d\sqrt{s_{\gamma\gamma}}}
  \hat{\sigma}_{\gamma \gamma \to \ell_i \bar\ell_j}(s_{\gamma\gamma})
  \equiv{\cal L}_{e^{+}e^{-}}\sigma_{\gamma \gamma \to \ell_{i} \bar\ell_{j}}(s)~,
\end{eqnarray}
where $d{\cal L}_{\gamma\gamma}$/$d\sqrt{s}_{\gamma\gamma}$ denotes as the
photon-beam luminosity distribution; $\sigma_{\gamma \gamma \to
\ell_i \bar\ell_j}(s_{ee})$, where $s_{ee}$ is the squared center-of-mass energy of
$e^{+}e^{-}$ collision, is the effective cross section of
$\gamma \gamma \to \ell_{i} \bar\ell_{j}$ and in the optimum case it
can be written as \cite{photon collider}
\begin{eqnarray}
\sigma_{\gamma \gamma \to \ell_i \bar\ell_j}(s)&=&
  \int_{\sqrt{a}}^{x_{max}}2zdz\hat{\sigma}_{\gamma \gamma \to \ell_{i} \bar\ell_{j}}
  (s_{\gamma\gamma}=z^2s) \int_{z^{2/x_{max}}}^{x_{max}}\frac{dx}{x}
 F_{\gamma/e}(x)F_{\gamma/e}(\frac{z^{2}}{x})~.\label{cross}
\end{eqnarray}
Here $F_{\gamma/e}$ is the energy spectrum of the back-scattered photon for the
unpolarized initial electron and laser photon beams, which can be written as
\begin{eqnarray}
F_{\gamma/e}(x)&=&\frac{1}{D(\xi)}\left[1-x+\frac{1}{1-x}-\frac{4x}{\xi(1-x)}
  +\frac{4x^{2}}{\xi^{2}(1-x)^{2}}\right]
\end{eqnarray}
with
\begin{eqnarray}
D(\xi)&=&(1-\frac{4}{\xi}-\frac{8}{\xi^{2}})\ln(1+\xi)
  +\frac{1}{2}+\frac{8}{\xi}-\frac{1}{2(1+\xi)^{2}}.
\end{eqnarray}
The definitions of parameters
$\xi$  and $x_{max}$ can be found in Ref.\cite{photon
collider} and  we choose $\xi=4.8$ and $x_{max}=0.83$ in the numerical calculation.

As for the SM parameters involved, we take \cite{pdg}
\begin{eqnarray}
m_{\mu}=0.106{\rm ~GeV}, m_{\tau}=1.777{\rm ~GeV}, m_e=0.511{\rm ~MeV},
\alpha=1/128.8,~\sin^2\theta_W=0.223. \nonumber
\end{eqnarray}

As the neutrino masses are quite small, the triplet VEV $v_\Delta$ which is responsible for neutrino masses should also be small. Small triplet VEVS are possible and could even be natural
\cite{09043640} by adjusting various parameters in the most general form of the Higgs potential.
There are two possible realization, firstly, authors of Ref. \cite{0509152}  point out that the lepton number is explicitly violated at very low energy scale $M_S$, which will
result in a tiny $v_\Delta$.  Secondly, even if the energy scale $M_S$ is not so tiny, i.e, $M_S \sim v$ ($v=246$ GeV ) $v_\Delta$ can be naturally small, which is denoted as a "type II seesaw mechanism" \cite{09043640}.
In our work, we will choose the VEV of the triplet $v_\Delta$ at the order of the typical neutrino mass upper limit, i.e, $v_\Delta \sim 1$ eV.

For the charged Higgs masses, the constraints are quite loose. Rough estimation can be obtained by the fact that the Higgs bosons that compose $\Delta$ will obtain masses at the electroweak scale \cite{masiero,0509152,09043640,type2-roadmap}, with
a neutral CP-even Higgs bosons playing the role of the standard Higgs with a mass at about $125$GeV \cite{atlas,cms}.
So we take the masses of the scalars other than the standard Higgs boson to lie in the range of a few of hundred GeV.
We assume the masses degenerate unless with otherwise statement, i.e, $m_{\phi^{\pm\pm}}$ = $m_{\phi^\pm}$ =  $m_\phi$.

Limits on the scalar mass $m_\phi$ can also be obtained by studying its effects on
various lepton flavor violating (LFV) constraints\cite{9511297,09043640,07124019}.
It is too weak and does not conflict with the assumption that the scalar masses $m_\phi$
are in the order of hundred GeV. We assume that the scalar mass $m_\phi$ is less than $1$ TeV. To investigate the
dependence of the cross sections on it, three classical values: $m_\phi=200,~500,~1000$ GeV are
taken in our calculations.

 The Maki-Nakagawa-Sakata (MNS) matrix $V_\MNS$
diagonalize the neutrino mass matrix mass can be written as \cite{Maki:1962mu,07124019} :
\begin{equation}
V_\MNS^{} =
\bmaT
c_{12}c_{13}                        & s_{12}c_{13}                  & s_{13}e^{-i\delta} \\
-s_{12}c_{23}-c_{12}s_{23}s_{13}e^{i\delta}  & c_{12}c_{23}-s_{12}s_{23}s_{13}e^{i\delta}  & s_{23}c_{13} \\
s_{12}s_{23}-c_{12}c_{23}s_{13}e^{i\delta}   & -c_{12}s_{23}-s_{12}c_{23}s_{13}e^{i\delta} & c_{23}c_{13}
\emaT
\,,
\end{equation}
Here $s_{ij}\equiv\sin\theta_{ij}$ and $c_{ij}\equiv \cos\theta_{ij}$; $\delta$ denotes the CP-phase.
 For majorana neutrinos, two additional phases should be added and then the mixing matrix $V$ is changed into
\begin{eqnarray}
 V = V_\MNS \times
     \text{diag}( 1, e^{i\phi_1 /2}, e^{i\phi_2 /2}),
\end{eqnarray}
where the discussions of the majorana phases $\phi_1$ and $\phi_2$ can be found
in Ref.~\cite{Schechter:1980gr, Mphase,07124019}.

In the HTM the triplet Yukawa coupling $h_{ij}$ is directly connected to
the neutrino mass matrix ($m_{ij}$), just as shown in Eq. (\ref{nu_mass}), which is
the phenomenologically attractive feature of this model.
Actually, the Eq.~(\ref{nu_mass}) can be rewritten in the basis of the three diagonal Dirac neutrino masses
by the MNS (Maki-Nakagawa-Sakata) matrix $V_\MNS$~ \cite{Maki:1962mu,09043640},
\begin{equation}
h_{ij}=\frac{m_{ij}}{\sqrt{2}v_\Delta}=\frac{1}{\sqrt{2}v_\Delta}
\left[
 V_\MNS
 \text{diag}(m_1,m_2 e^{i\phi_1},m_3 e^{i\phi_2})
 V_\MNS^T
\right]_{ij}
\label{hij}
\end{equation}

By expanding Eq. (\ref{hij}), the explicit expressions of $h_{ij}$  can be found \cite{Garayoa:2007fw, 07124019, Kadastik:2007yd, Perez:2008ha,09043640}:
\begin{eqnarray}
h_{ee} &=& \frac{1}{\sqrt{2} v_\Delta}
 \Bigl(
  m_1 c_{12}^2 c_{13}^2
  + m_2 s_{12}^2 c_{13}^2 e^{i\phi_1}
  + m_3 s_{13}^2 e^{-2i\delta} e^{i\phi_2}
 \Bigr)\,,
 \nonumber \\
h_{e\mu} &=& \frac{1}{\sqrt{2} v_\Delta}
 \Bigl\{
  m_1 ( -s_{12}c_{23} - c_{12}s_{23}s_{13} e^{i\delta} ) c_{12}c_{13}
 \nonumber \\
 &&\hspace{30mm}
 {}+ m_2 ( c_{12}c_{23} - s_{12}s_{23}s_{13} e^{i\delta} ) s_{12}c_{13} e^{i\phi_1}
  + m_3 s_{23}c_{13}s_{13} e^{-i\delta} e^{i\phi_2}
 \Bigr\}\,,
 \nonumber \\
h_{e\tau} &=& \frac{1}{\sqrt{2} v_\Delta}
 \Bigl\{
  m_1 ( s_{12}s_{23} - c_{12}c_{23}s_{13} e^{i\delta} ) c_{12}c_{13}
 \nonumber \\
 &&\hspace{30mm}
  {}+ m_2 ( -c_{12}s_{23} - s_{12}c_{23}s_{13} e^{i\delta} ) s_{12}c_{13} e^{i\phi_1}
  + m_3 c_{23}c_{13}s_{13} e^{-i\delta} e^{i\phi_2}
 \Bigr\}\,,
  \nonumber \\
h_{\mu\mu} &=& \frac{1}{\sqrt{2} v_\Delta}
 \Bigl\{
  m_1 ( -s_{12}c_{23} - c_{12}s_{23}s_{13} e^{i\delta} )^2
  + m_2 ( c_{12}c_{23} - s_{12}s_{23}s_{13} e^{i\delta} )^2 e^{i\phi_1}
  + m_3 s_{23}^2 c_{13}^2 e^{i\phi_2}
 \Bigr\}\,,
  \nonumber \\
h_{\mu\tau} &=& \frac{1}{\sqrt{2} v_\Delta}
 \Bigl\{
  m_1 ( -s_{12}c_{23} - c_{12}s_{23}s_{13} e^{i\delta} )
        ( s_{12}s_{23} - c_{12}c_{23}s_{13} e^{i\delta} )
 \nonumber \\
 &&\hspace{20mm}
  {}+ m_2 ( c_{12}c_{23} - s_{12}s_{23}s_{13} e^{i\delta} )
        ( -c_{12}s_{23} - s_{12}c_{23}s_{13} e^{i\delta} ) e^{i\phi_1}
  + m_3 c_{23}s_{23} c_{13}^2 e^{i\phi_2}
 \Bigr\}\,,
  \nonumber \\
h_{\tau\tau} &=& \frac{1}{\sqrt{2} v_\Delta}
 \Bigl\{
  m_1 ( s_{12}s_{23} - c_{12}c_{23}s_{13} e^{i\delta} )^2
  + m_2 ( -c_{12}s_{23} - s_{12}c_{23}s_{13} e^{i\delta} )^2 e^{i\phi_1}
  + m_3 c_{23}^2 c_{13}^2 e^{i\phi_2}
 \Bigr\}\,.
 \nonumber \\
\label{hij_expressions}
\end{eqnarray}
From above equation, we can see that the couplings $h_{ij}$ depend on the
following nine parameters: the mass-squared differences $\Delta m^2_{21}$, $\Delta m^2_{31}$,
the mass of the lightest neutrino $m_0$,
 three mixing angles $\theta_{12}$, $\theta_{13}$, $\theta_{23}$,
 and three complex phases ($\delta$, $\phi_1$, $\phi_2$).

Different neutrino oscillation experiments, such as the solar~\cite{solar}, atmospheric~\cite{atm}, accelerator~\cite{acc}, and reactor neutrinos~\cite{daya-bay}, can be used to determine the mass-squared
differences ($\Delta m^2_{21}$, $\Delta m^2_{31}$) and the mixing angles ($\theta_{12}$, $\theta_{13}$, $\theta_{23}$).
The preferred values are given in the following:
\begin{eqnarray}
\Delta m^2_{21} \equiv m^2_2 -m^2_1
\simeq 7.9\times 10^{-5} {\rm eV}^2 \,,~~
|\Delta m^2_{31}|\equiv |m^2_3 -m^2_1|
\simeq 2.7\times 10^{-3} {\rm eV}^2\,, \\ \nonumber
\sin^22\theta_{12}\simeq 0.86 \,,~~~~ \sin^22\theta_{23}\simeq 1 \,,~~~~
\sin^22\theta_{13}\simeq 0.089\,.~~~~~~~~~~~~
\label{obs_para}
\end{eqnarray}

 Since the sign of $\Delta m_{31}^2$ is unknown for now,
there are two neutrino mass-hierarchy patterns. 
One possibility is the normal hierarchy (NH), with $\Delta m^2_{31} >0$ where $m_1 < m_2 < m_3$
and the other is the Inverted hierarchy (IH) with $\Delta m^2_{31} <0$
where $m_3 <  m_1 < m_2$\cite{09043640}.

It is very difficult, even impossible to some extent, to extract
informations on majorana phases solely
from the neutrino oscillation experiments~\cite{phase-0nbb,09043640}.
Therefore, it is worthwhile to
consider other possibilities, such as the
LFV processes in the context of HTM, to determine the majorana phases.

As mentioned in Ref. \cite{0304069,09043640,07124019},
one can define four cases of the majorana phases as follows:
Case~I $(\phi_1=0,\phi_2=0)$;
Case~II $(\phi_1=0,\phi_2=\pi)$;
Case~III $(\phi_1=\pi,\phi_2=0)$;
Case~IV $(\phi_1=\pi,\phi_2=\pi)$.
 In this work we will study in detail the
dependence of $\gamma\gamma \to \ell_i \bar\ell_j$ in each case with
the new values of $\theta_{13}$ given
by Daya Bay\cite{daya-bay}, i.e, $sin ^22\theta_{13}\simeq 0.089$.

\section{Numerical Results and Discussions}
The lepton flavor changing production processes $\gamma\gamma \to \ell_i \bar\ell_j$, including the $e\mu$,
$e\tau$ and  $\mu\tau$, can be different in normal hierarchy case and inverted hierarchy case, respectively.  We will discuss such two possibilities with the choices of majorana phase from case-I to case-IV.

 The relevant parameters in this process include the neutrino parameters, the scalar masses and the scale parameter.
The neutrino parameters are: $\Delta m^2_{21}$, $\Delta m^2_{31}$, $m_0$, $\theta_{12}$, $\theta_{13}$, $\theta_{23}$,
 and $\delta$, $\phi_1$, $\phi_2$. $\Delta m^2_{21}$, $\Delta m^2_{31}$. The parameters $\theta_{12}$, $\theta_{13}$, $\theta_{23}$,
 and $\delta$ take the values given by experiments in Eq. (\ref{obs_para}) and we take $\delta=0$. Four different choices of $\phi_1$, $\phi_2$ (case-I to case-IV) will be discussed in our study. The remaining $m_0$, the lightest neutrino mass, is quite small. The upper limit for the summation of all the neutrino masses \cite{09115291} is given by
$\sum m_\nu \leq 0.28 $ eV ($95\%~CL $) assuming a flat
$\Lambda$CDM cosmology. Thus we will take the mass range of  $m_0$ to be $0\leq m_0 \leq 0.3$ eV as an estimation..

It has been shown that the coupling constants $h_{ij}$, especially $h_{ee},~h_{e\mu}$ which are the functions
 of the neutrino flavor parameters, should satisfy
certain constraint, i.e. $h_{ee},~h_{e\mu} \sim 0$,  $h_{e\tau},~h_{\tau\mu} < 1$  \cite{09094943,09043640,9511297,0304254,0304069}.
We will take into account such constraints in the results we obtained.

Finally, the charge conjugate process $\gamma\gamma \to \bar \ell_i \ell_j$ production channel
will also be included in our numerical study.

\subsection{ Normal Hierarchy}
By simple estimations from the expressions of the $h_{ij}$($i,j=e,~\mu,~\tau$),
 we can see that their values have large hierarchy. So we can discuss the
 productions $\gamma\gamma \to \mu\bar e$, 
 $\tau\bar e$, 
 and $ \mu\bar \tau$  
 one by one in the four cases of $\phi_1$ and $\phi_2$.
\subsubsection{$\gamma\gamma \to \mu\bar e$ }   
\def\figsubcap#1{\par\noindent\centering\footnotesize(#1)}
\begin{figure}[bht]%
\begin{center}
\hspace{-2.5cm}
 \parbox{7.05cm}{\epsfig{figure=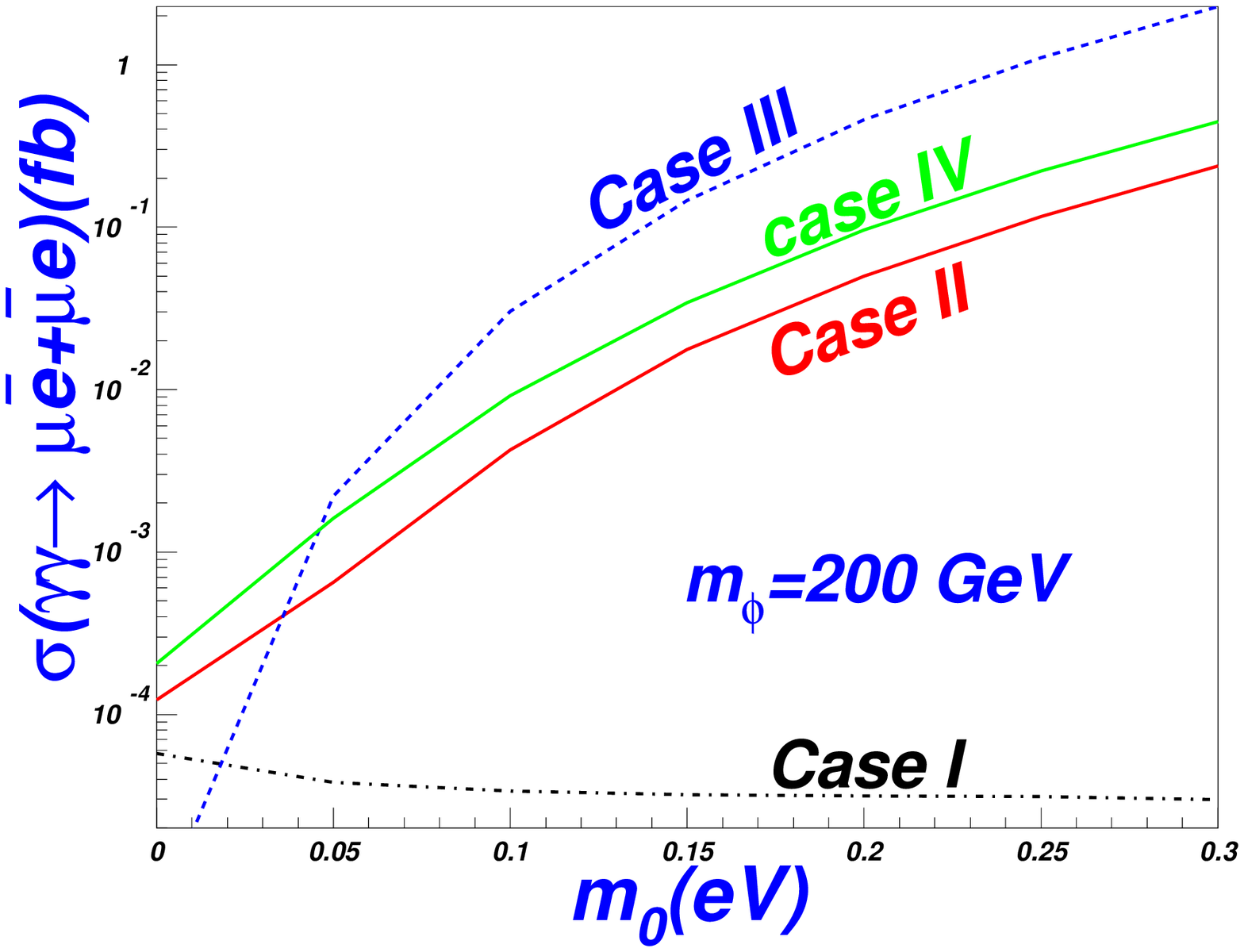,width=7.25cm} \figsubcap{a} }
 \parbox{7.05cm}{\epsfig{figure=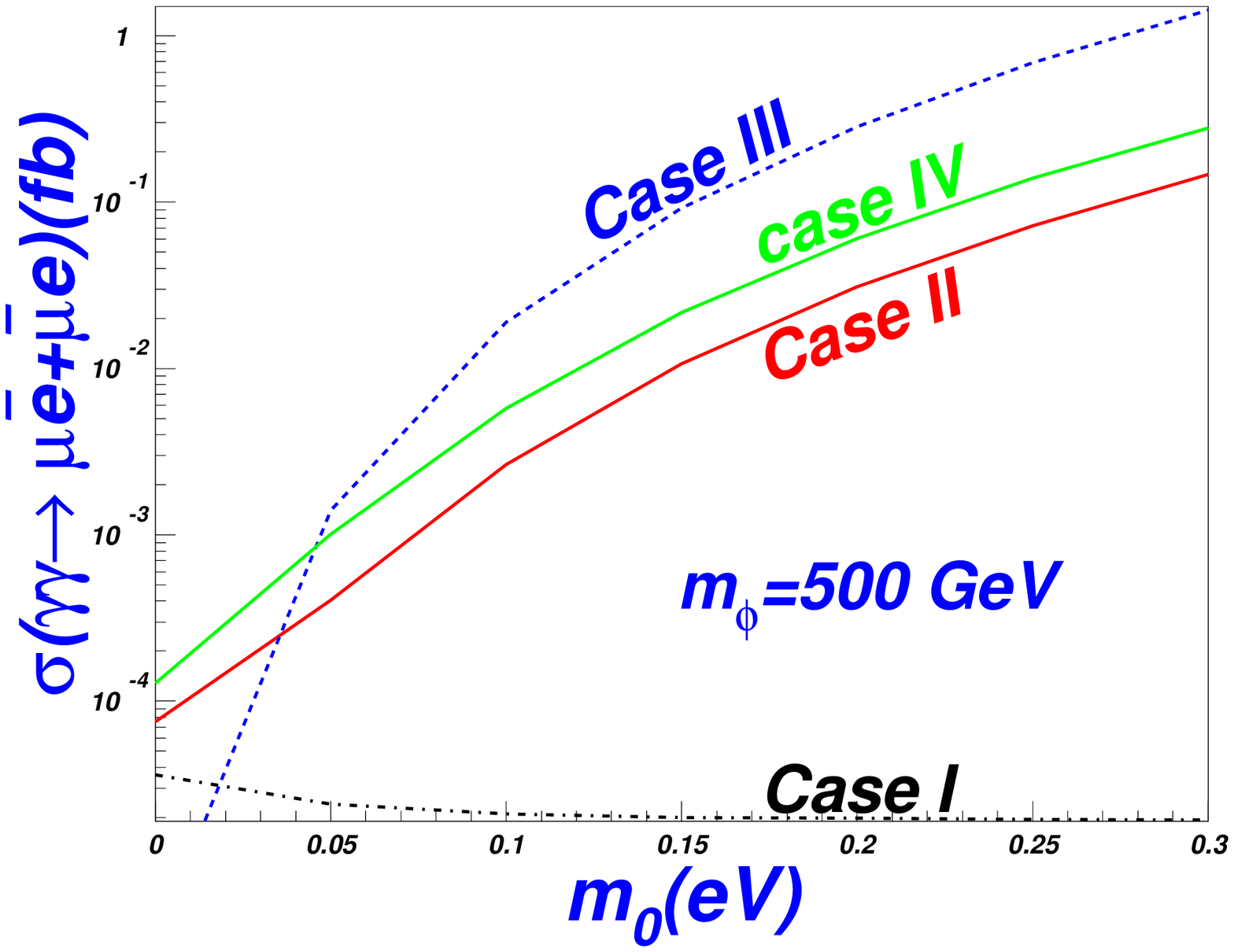,width=7.25cm} \figsubcap{b} }
 \parbox{7.05cm}{\epsfig{figure=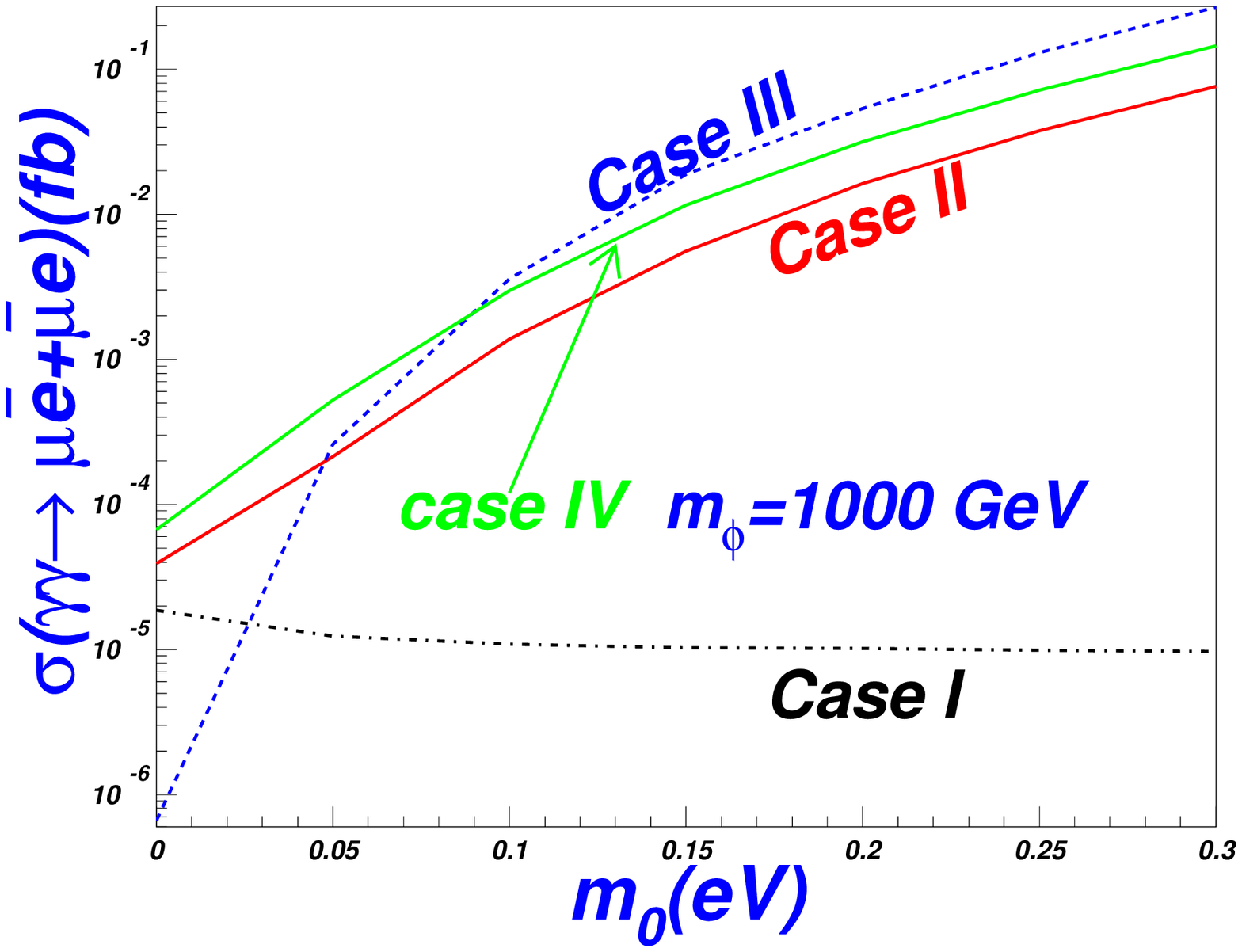,width=7.25cm} \figsubcap{c} }
  \caption{ The cross section $\sigma$ of the LFV process $\gamma\gamma
            \to  \mu \bar e $ as a function of the neutrino mass $m_0$
             for  Case I to Case IV 
             with different scalar mass: $m_\phi=200,~500,~1000$ GeV
      for $\sqrt{s}=500$ GeV. 
\label{fig2} }
\end{center}
\end{figure}
Figure \ref{fig2} shows the cross sections of the $\gamma\gamma \to \mu\bar e$,
 varying with respect to the lightest neutrino mass $m_0$, with different scalar mass $m_\phi=200,~500,~1000$ GeV. From the figure, we can see that the production rates
 increase with the increasing $m_0$. When $m_0$ is small, for example, less than $0.1$ eV, the cross sections are less than $0.01$ fb in most of parameter space.
 But when $m_0$ becomes large, the production rates may arrive at $2$ fb in the optimum region of case III.
 We can also see that the cross section is also affected by the scalar mass, which may vary from $200$ GeV to $1000$ GeV.
Such influence is however much more smaller than that from $m_0$. So in the latter discussion, we will take $m_\phi=200 $ GeV.

\begin{figure}[bht]%
\begin{center}
\hspace{-2.5cm}
 \parbox{9.05cm}{\epsfig{figure=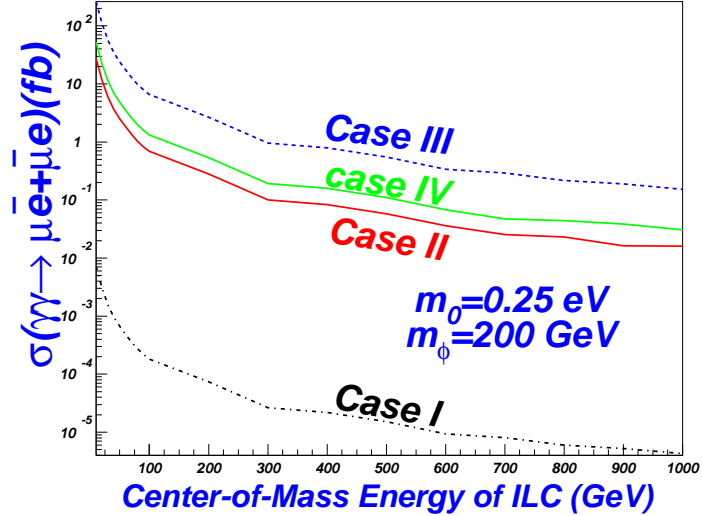,width=9.25cm} } 
 \caption{ The cross section $\sigma$ of the LFV process $\gamma\gamma
            \to   \mu \bar e $ as a function of the center-of-mass $\sqrt{s}$ 
                for  Case I to Case IV, 
                with the scalar mass $m_\phi=200$ GeV and the minimal neutrino mass $m_0=0.25$ eV.  
\label{fig3} }
\end{center}
\end{figure}

The center-of-mass dependence of the process $\gamma\gamma \to \mu\bar e$ is displayed in figure \ref{fig3}, with
 $m_0 =0.25$ eV and $m_\phi=200$ GeV. From figure \ref{fig3}, we can see that the production can be much larger when the $\sqrt{s}$ is small. For example, when $\sqrt{s} =10$ GeV, the production rate can arrive at $260$ fb in case III. But for larger center-of-mass energy, the cross sections will become small. For example, the cross section arrives at $0.55$ fb when $\sqrt{s} =500$ GeV.

In Case I, II and IV, the cross sections are a bit smaller than that of Case III. In Case I with $\phi_1=\phi_2=0$, the cross section is at the order of the $10^{-3}$ fb and quite small. But in Case II and IV, the cross sections can arrive at tens of fb, though smaller than that of Case III.

 We can see in Figure \ref{fig3} that the production rates of the process $\gamma\gamma \to \mu\bar e$ decrease with the increasing center-of-mass energy $\sqrt{s}$, which is reasonable since there is no s-channel charged scalars contributions to the lepton flavor changing process and the large masses of the inner line particles may suppress the production rates further. Similar behaviors are also shown in some supersymmetric models \cite{susy-r-con1,rrmutau-susy}.
\def\figsubcap#1{\par\noindent\centering\footnotesize(#1)}

\subsubsection{$\gamma\gamma \to \bar e\tau$ and $\gamma\gamma \to \bar  \mu\tau$ }
\def\figsubcap#1{\par\noindent\centering\footnotesize(#1)}
\begin{figure}[bht]%
\begin{center}
\hspace{-2.5cm}
 \parbox{7.05cm}{\epsfig{figure=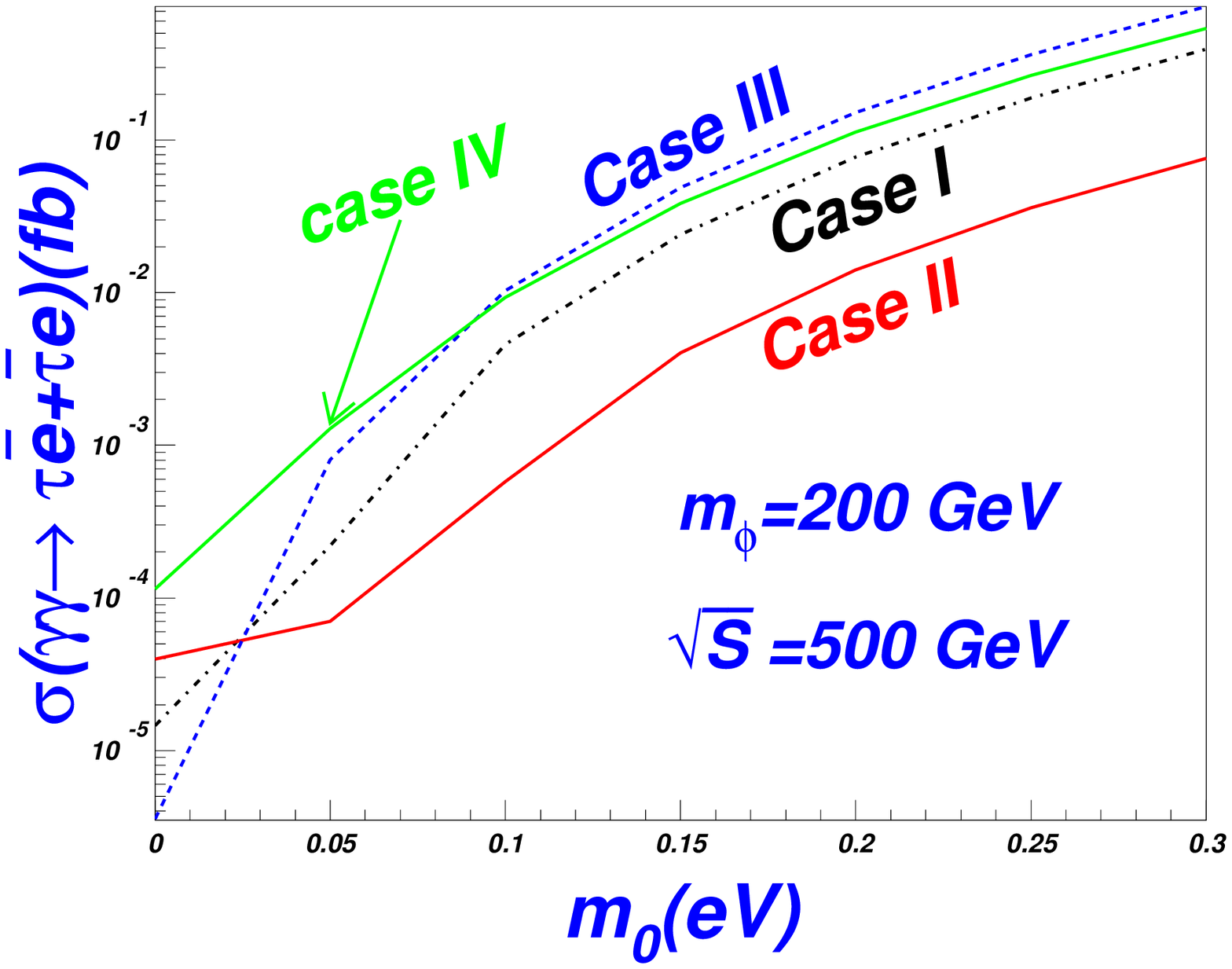,width=7.25cm} \figsubcap{a} }
 \parbox{7.05cm}{\epsfig{figure=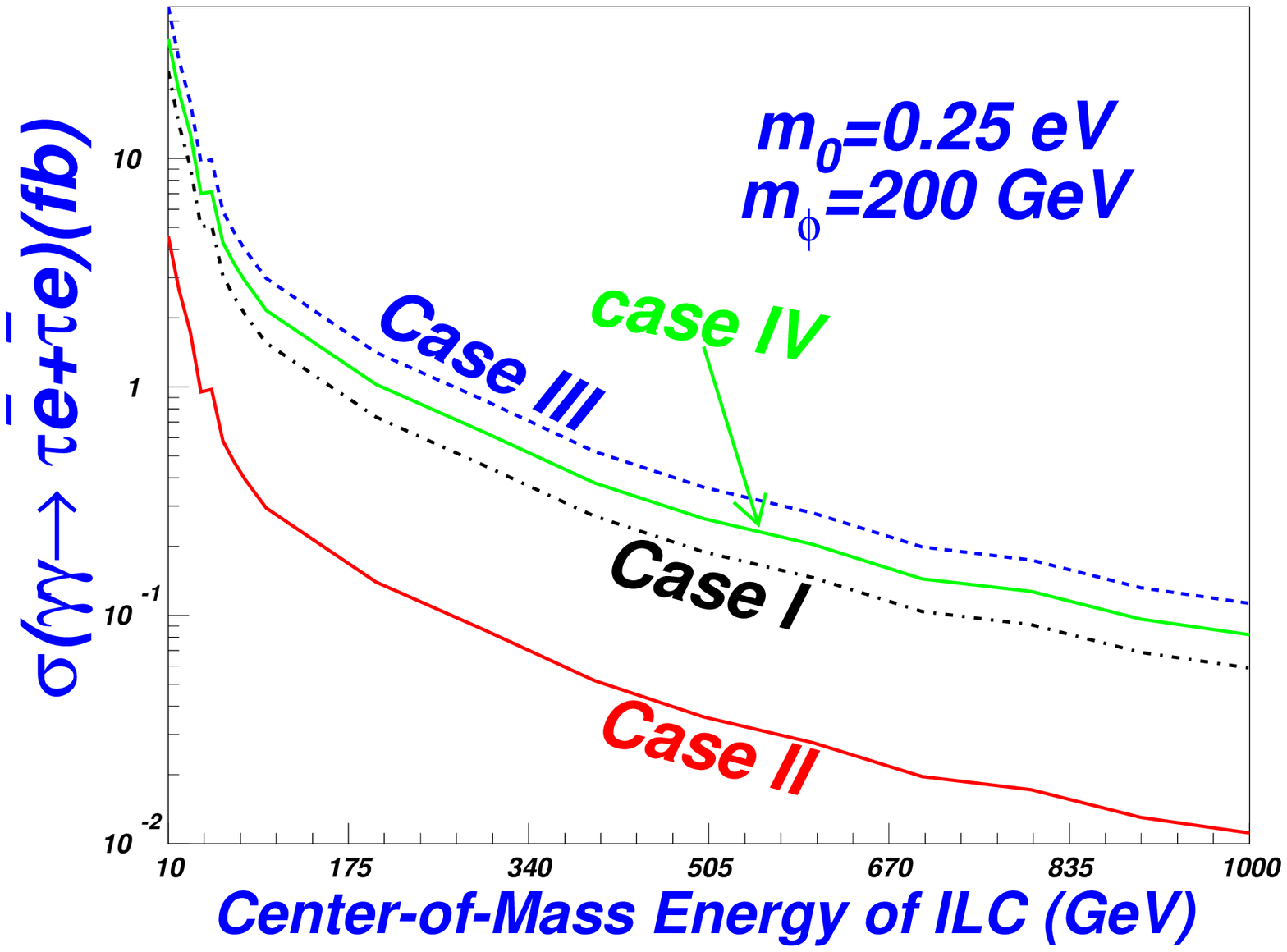,width=7.25cm} \figsubcap{b} }
 \caption{ The cross section $\sigma$ of the LFV process $\gamma\gamma
            \to   \tau \bar e $ and 
            as a function of the  minimal neutrino mass $m_0$ (a) (for $E=500$ GeV) and the center-of-mass energy E (b) (for  $m_0=0.25$ eV)  from  Case I to Case IV, with $m_\phi=200$ GeV. 
\label{fig4} }
\end{center}
\end{figure}
\def\figsubcap#1{\par\noindent\centering\footnotesize(#1)}
\begin{figure}[bht]%
\begin{center}
\hspace{-2.5cm}
 \parbox{7.25cm}{\epsfig{figure=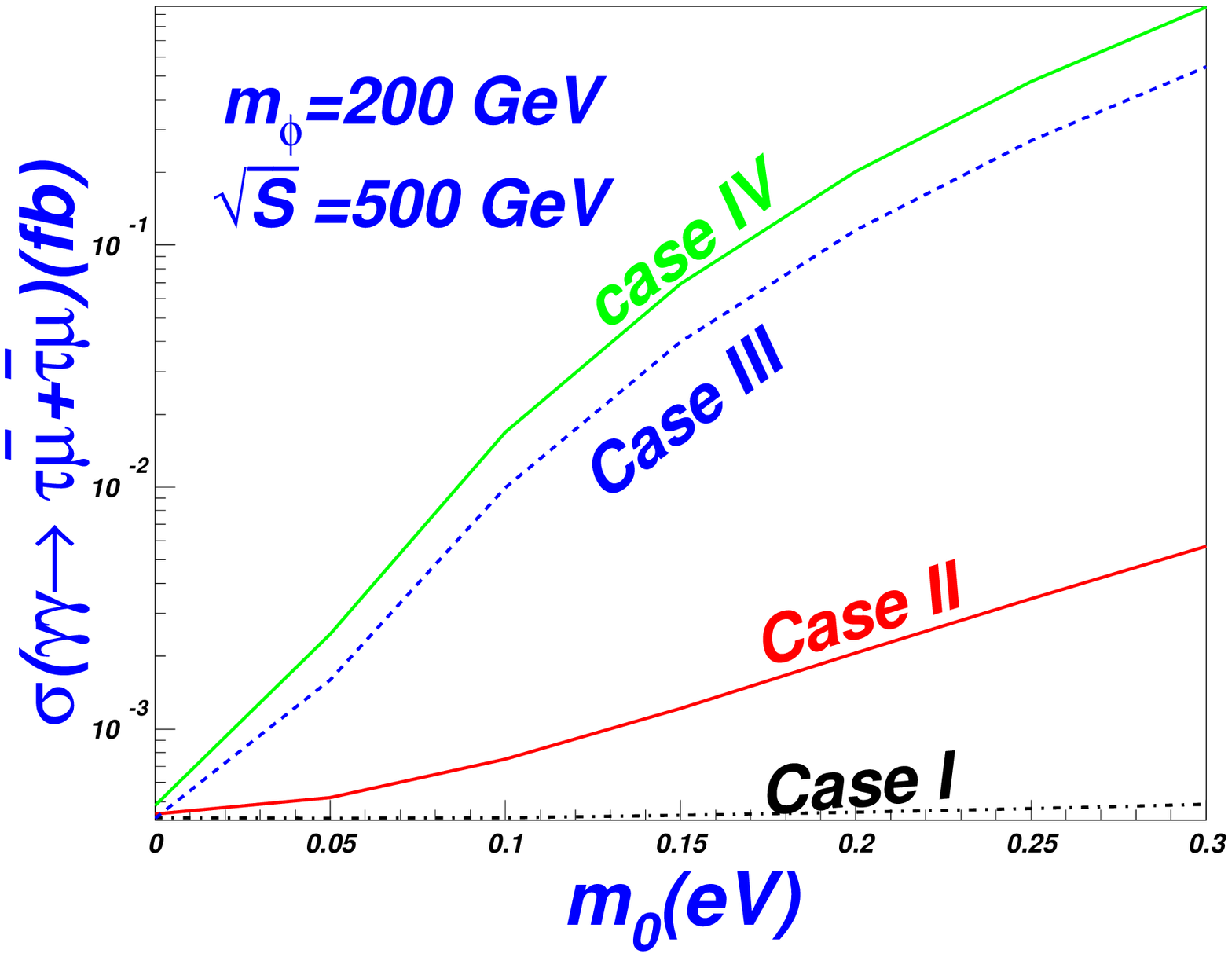,width=7.25cm} \figsubcap{a} }
 \parbox{7.25cm}{\epsfig{figure=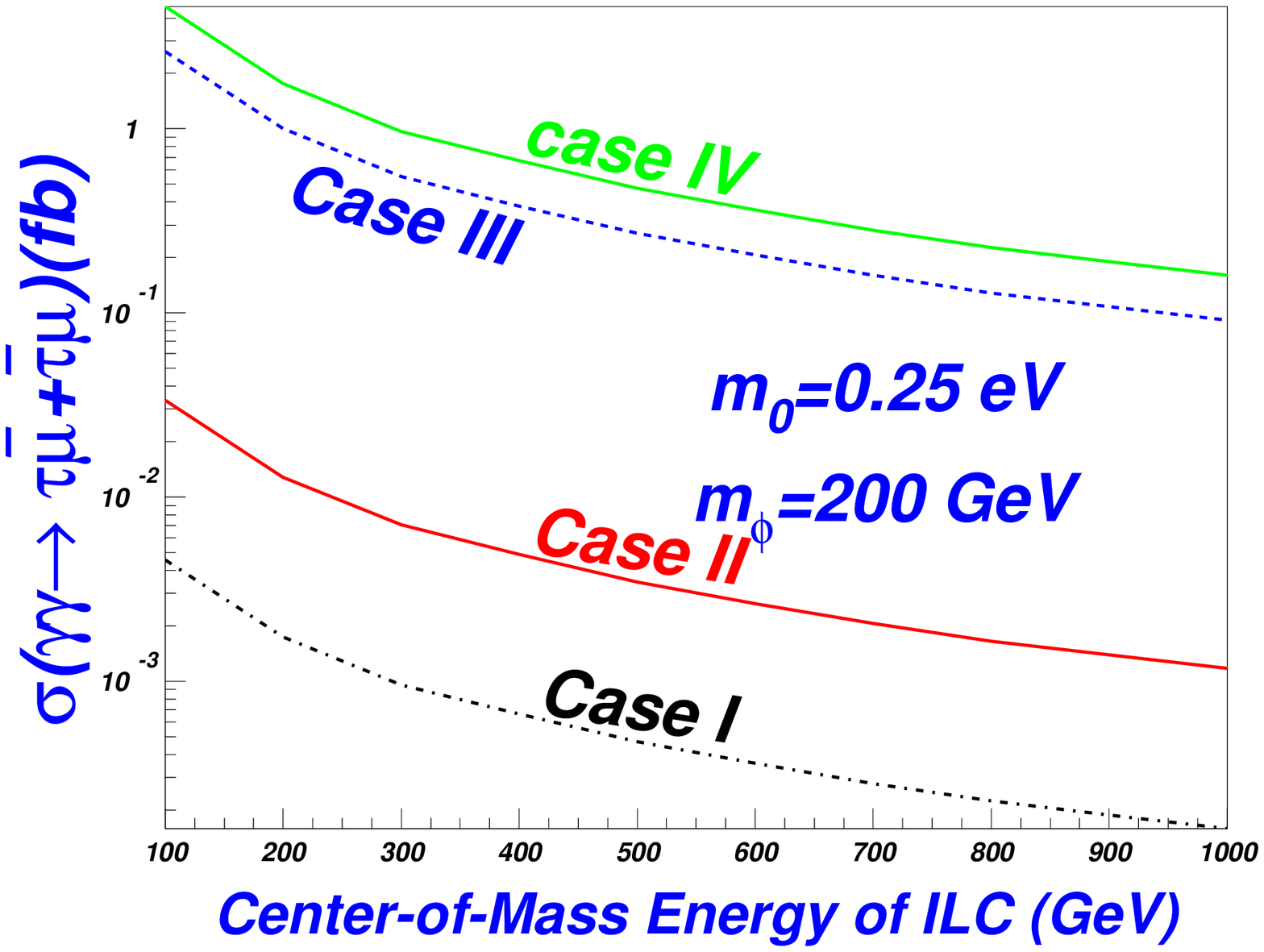,width=7.25cm} \figsubcap{b} }
 \caption{ Same as Figure  \ref{fig4}, but for $\gamma\gamma
            \to   \tau \bar \mu $.
\label{fig5} }
\end{center}
\end{figure}
 Figure \ref{fig4} and Figure \ref{fig5} show the cross sections of the $ \bar e \tau$ and $ \bar \mu \tau$ production in the $\gamma\gamma$ collision, varying with respect to the lightest neutrino mass $m_0$ and center-of-mass $\sqrt{s}$, with the scalar mass $m_\phi=200$ GeV for Case I, II, III and IV, respectively. We can see from them that both production rates are almost in the same order as the process $\gamma\gamma\to \mu \bar e$.

Figure  \ref{fig4} (a) and Figure  \ref{fig5} (a) show the $m_0$ dependence of the cross sections of the two processes $\gamma\gamma \to \bar e\tau$ and $\gamma\gamma \to \bar  \mu\tau$.  We can see that the production rates  increase with the increasing neutrino mass and decline with the raising center-of-mass energy of ILC.

Figure  \ref{fig4} (b) and Figure  \ref{fig5} (b) give the center-of-mass dependence of the
cross sections of the $ \bar e \tau$ and $ \bar \mu \tau$  production, from which we can see
that the behaviors are the same as those of the process $\gamma\gamma \to \mu \bar e $. Their production rates become smaller with increasing center-of-mass energy of the ILC.

We have seen from Figure  \ref{fig2}, \ref{fig4}, \ref{fig5} that the production rates of the lepton flavor changing processes increase with increasing $m_0$. This is justified since the flavor couplings are directly connected to the neutrino masses. So these processes may provide a good environment to detect the neutrino masses.

We can also see that the cross sections of the production processes $\bar e \mu$, $\bar e\tau$ and $\bar \mu\tau$ are almost in the same order with fixed center-of-mass energy of the ILC. For Case IV with $\phi=\pi,~\phi=\pi$ and setting $m_0=0.25$ eV and $\sqrt{s} = 500 $ GeV, the corresponding rates are $0.22$ fb, $0.264$ fb and $0.47$ fb, respectively.

\subsection{ the Inverted Neutrino Mass Hierarchy}
\def\figsubcap#1{\par\noindent\centering\footnotesize(#1)}
\begin{figure}[thb]%
\begin{center}
 \parbox{7.05cm}{\epsfig{figure=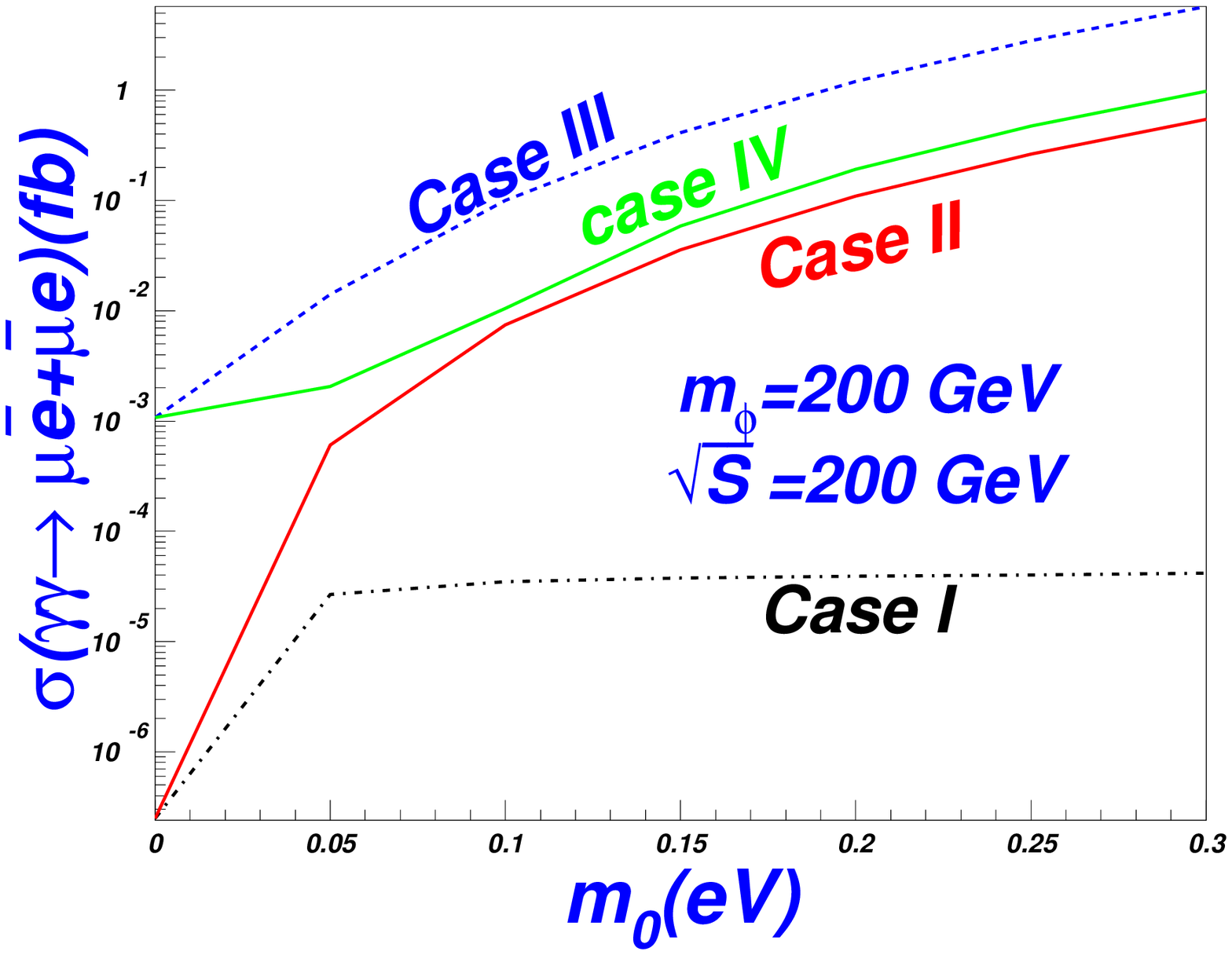,width=7.05cm} \figsubcap{a} }
 \parbox{7.05cm}{\epsfig{figure=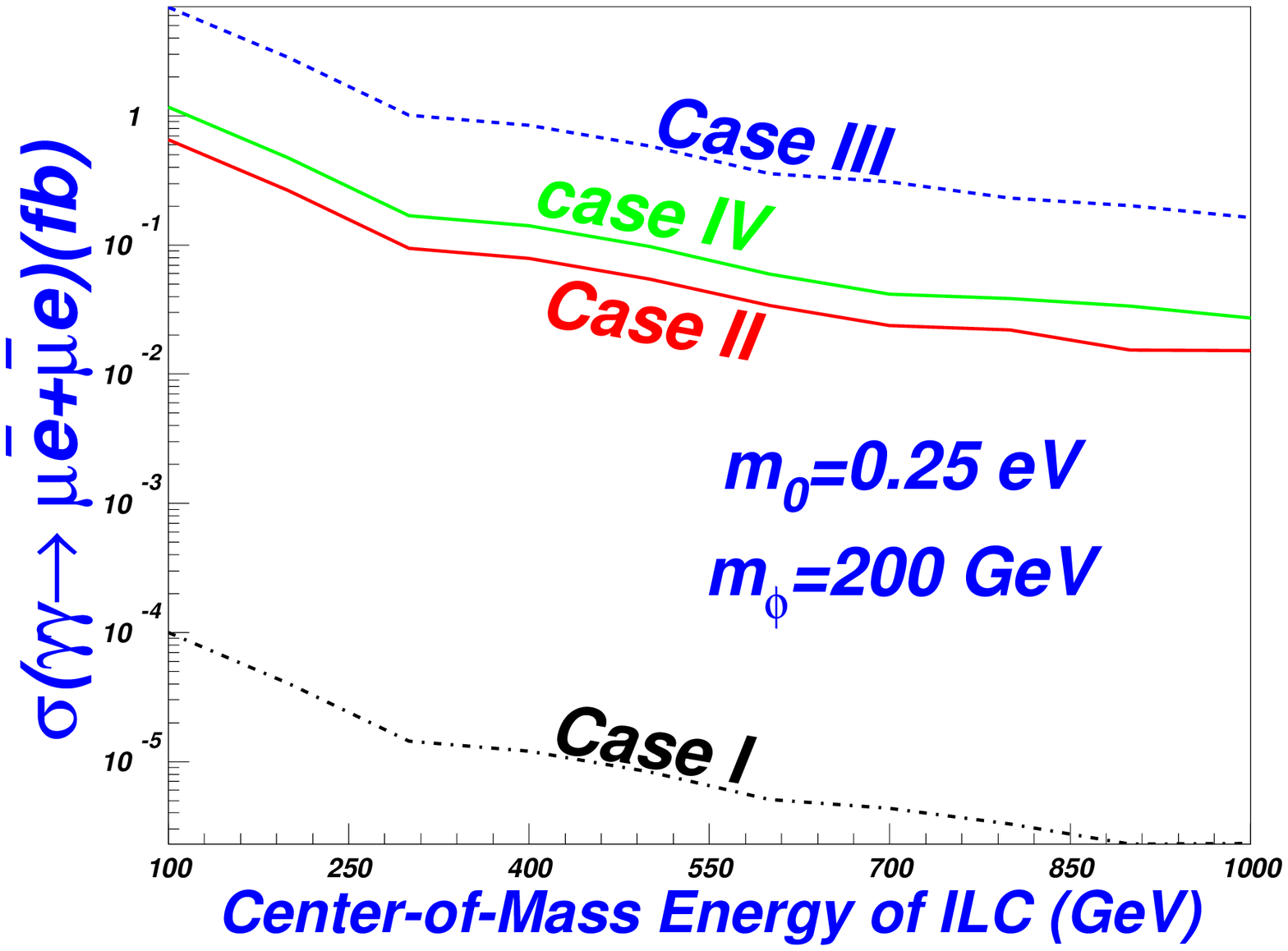,width=7.05cm} \figsubcap{b} }
 \caption{ The LFV process $\gamma\gamma \to   \mu \bar e$
 cross section $\sigma$ in the case of the Inverted Neutrino Mass Hierarchy
  as a function of the minimal neutrino mass $m_0$
and the center-of-mass energy of ILC 
 from  Case I to  Case IV: 
 with the scalar mass $m_\phi=200$ GeV. 
\label{fig6} }
\end{center}
\end{figure}
We also study the three productions in the inverted hierarchy case. The results can be found in the Figures  \ref{fig6},  \ref{fig7} and  \ref{fig8}, from which we can see that the production rates are almost the same as that in the normal hierarchy. So we will not discuss them in detail.

\def\figsubcap#1{\par\noindent\centering\footnotesize(#1)}
\begin{figure}[thb]%
\begin{center}
\parbox{7.05cm}{\epsfig{figure=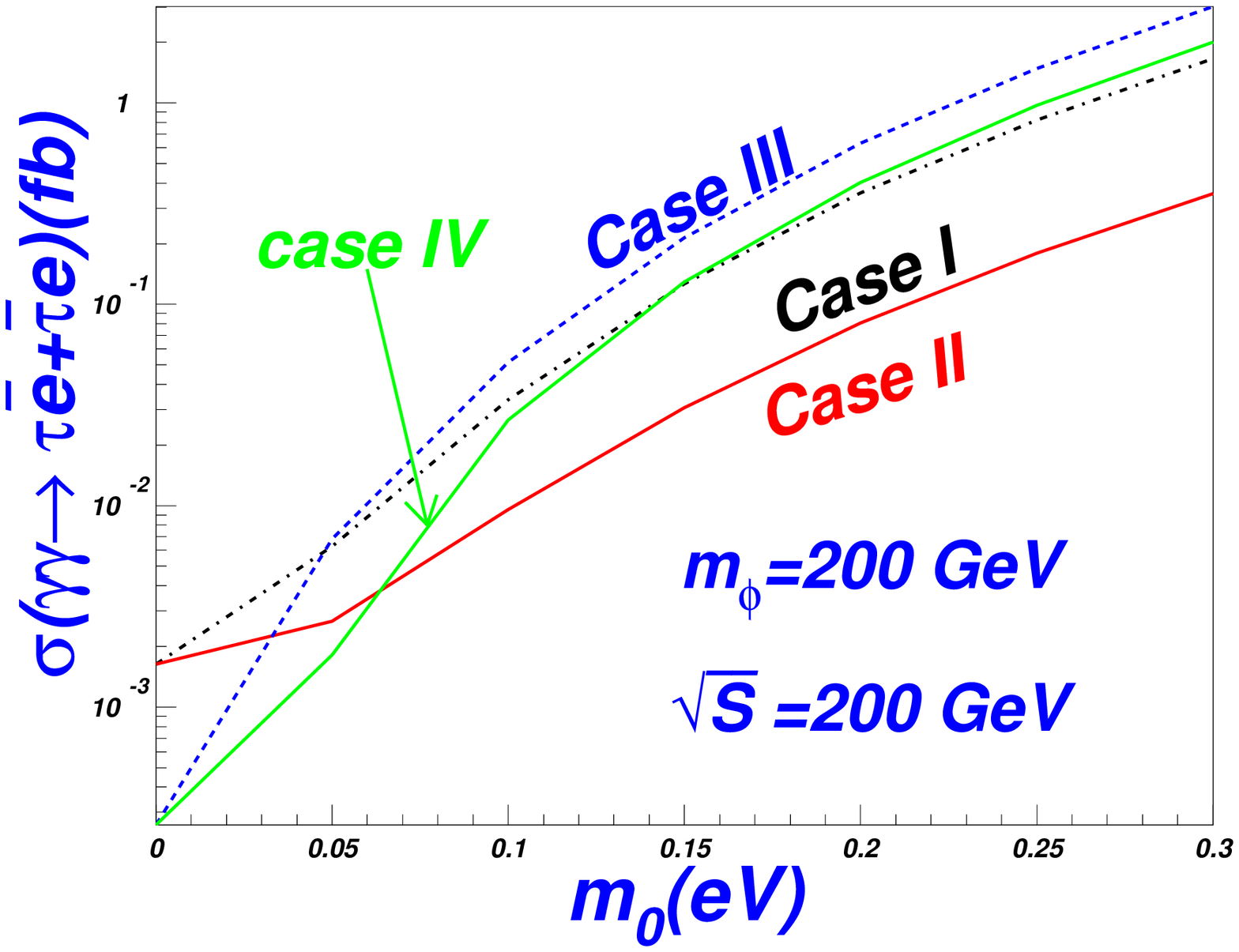,width=7.05cm} \figsubcap{a} }
\parbox{7.05cm}{\epsfig{figure=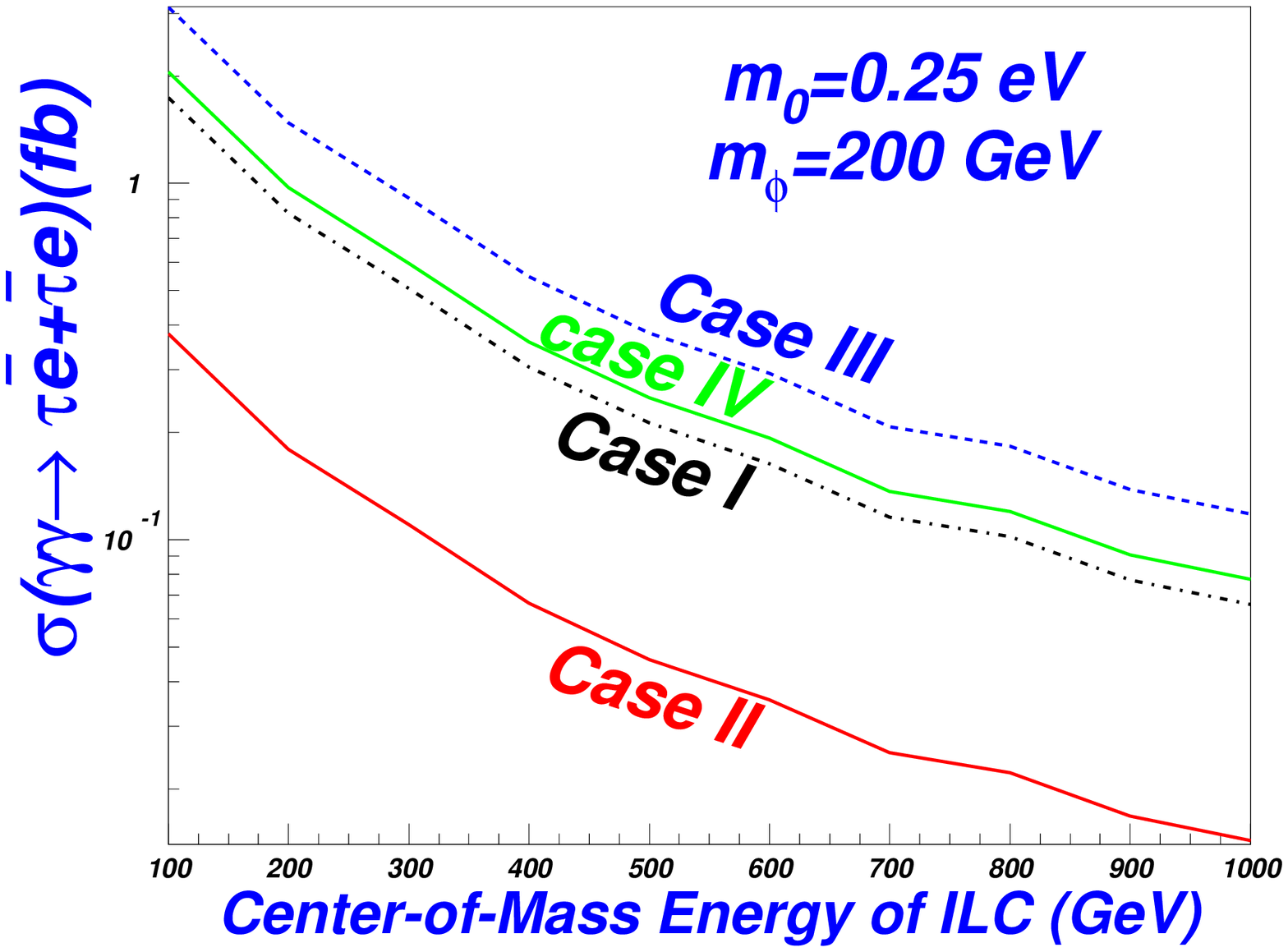,width=7.05cm} \figsubcap{b} }
\caption{ Same as Figure  \ref{fig6}, but for $\gamma\gamma \to   \tau \bar e$.
\label{fig7} }
\end{center}
\end{figure}

\def\figsubcap#1{\par\noindent\centering\footnotesize(#1)}
\begin{figure}[thb]%
\begin{center}
  \parbox{7.05cm}{\epsfig{figure=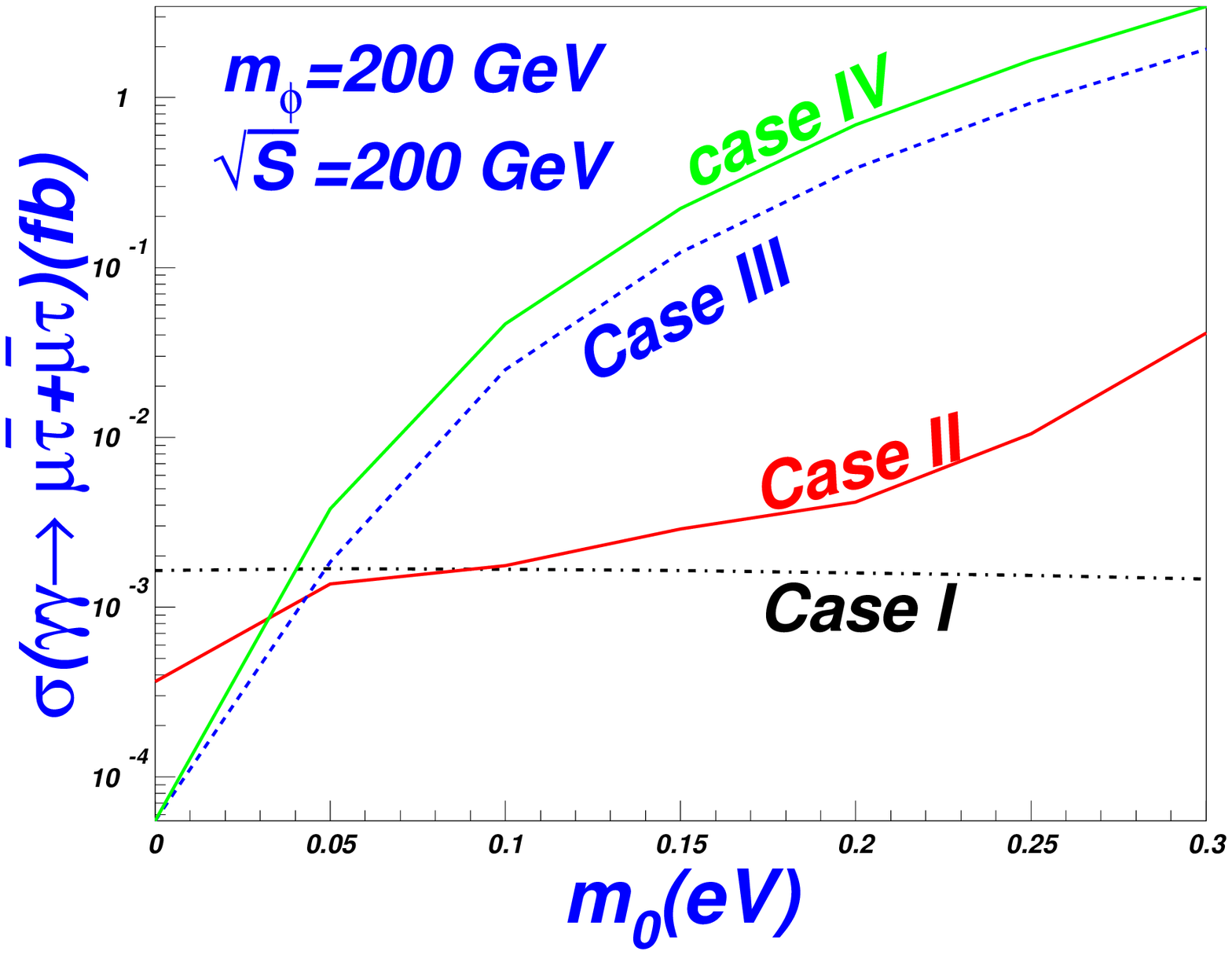,width=7.05cm} \figsubcap{a} }
  \parbox{7.05cm}{\epsfig{figure=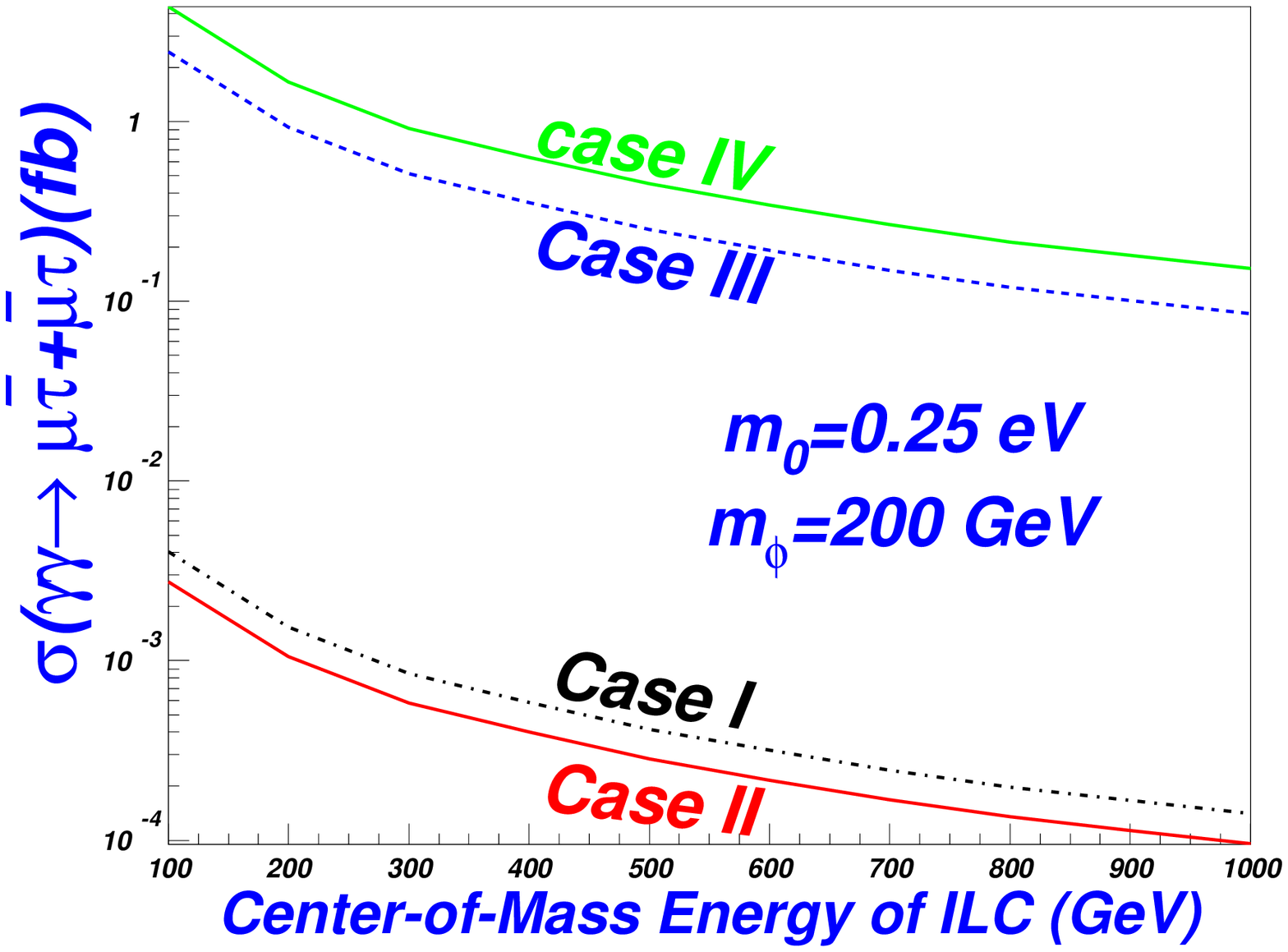,width=7.05cm} \figsubcap{b} }
 \caption{ Same as Figure  \ref{fig6}, but for $\gamma\gamma \to   \tau \bar \mu$.
\label{fig8} }
\end{center}
\end{figure}

\subsection{$h_{e\mu}= h_{ee}\approx 0$ constraints}
After the calculations of the cross sections for each of the four cases, we turn to the constraints from the
$\mu\to 3e$ and $\mu\to e\gamma$ \cite{09094943,09043640,9511297,0304254,0304069}
 and find that the couplings $h_{e\mu}$ and $h_{ee}$ should be quite small. We take the limit
 $h_{e\mu}= h_{ee}\approx 0$ to see the differences in the production rates.
We will discuss how the constraint $h_{e\mu}= h_{ee}\approx 0$ can affect the production rates with $h_{e\tau},~h_{\mu\mu},~h_{\mu\tau},~h_{\tau\tau}$ being fixed by the assumptions in Case I to Case IV.

Figure  \ref{fig9} gives the cross sections of the $\gamma\gamma \to e\mu,~ e\tau,~ \mu\tau$ production rates from Case I to Case IV, from which we can see that the production rates are different from those without the constraints.
To estimate the effects, we can compare the figure with Figures  \ref{fig2} (a), \ref{fig4} (a) and Figure  \ref{fig5} (a),
We find that the rates with the constraints  $h_{e\mu}= h_{ee}\approx 0$ are about one order lower than those without the constraints.
This is reasonable because the constraints $h_{e\mu}= h_{ee}\approx 0$
switch off the contributions from the Yukawa couplings $ h_{e\mu}$ and $ h_{ee}$. 
\def\figsubcap#1{\par\noindent\centering\footnotesize(#1)}
\begin{figure}[bht]%
\begin{center}
\hspace{-2.5cm}
 \parbox{7.05cm}{\epsfig{figure=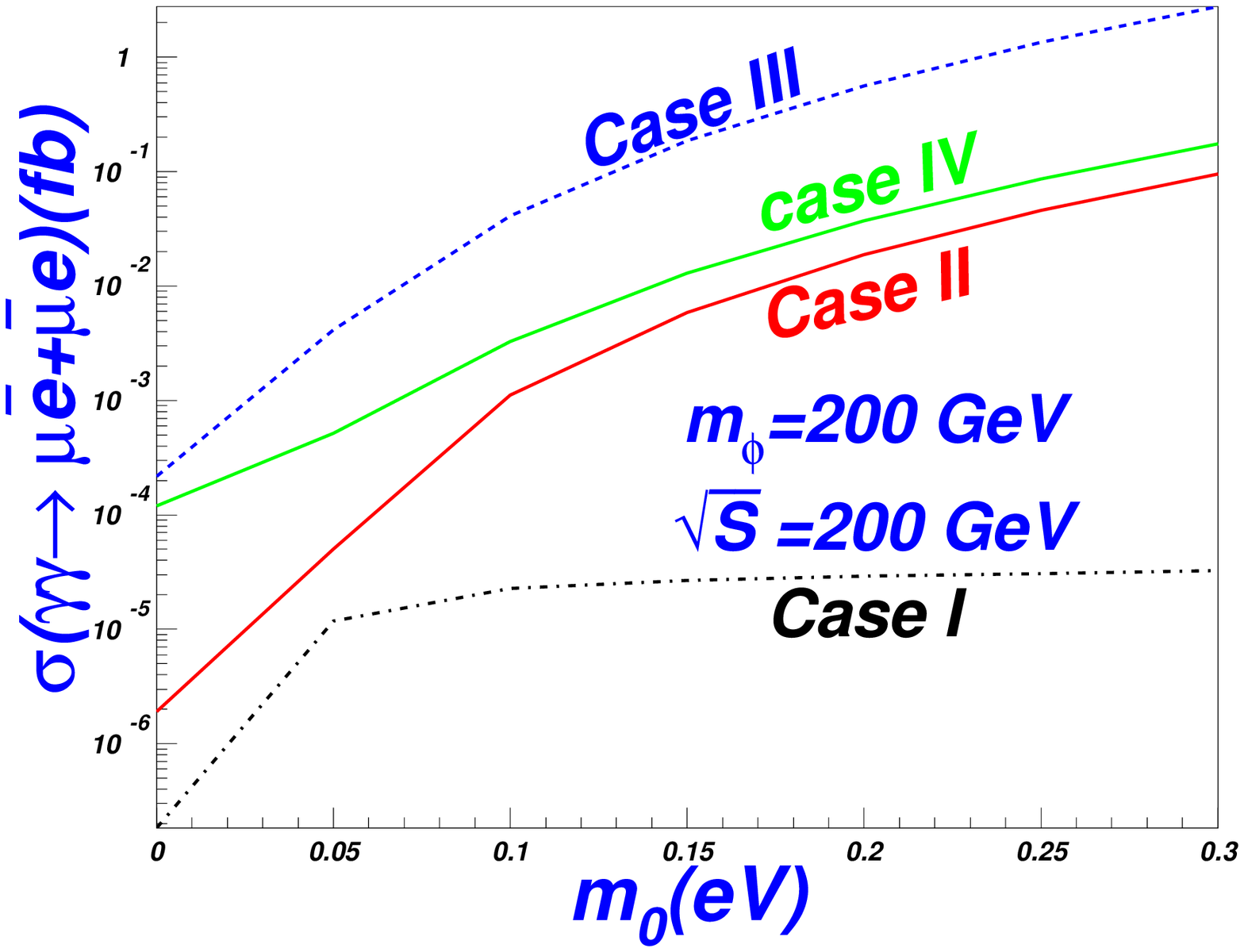,width=7.25cm} \figsubcap{a} }
 \parbox{7.05cm}{\epsfig{figure=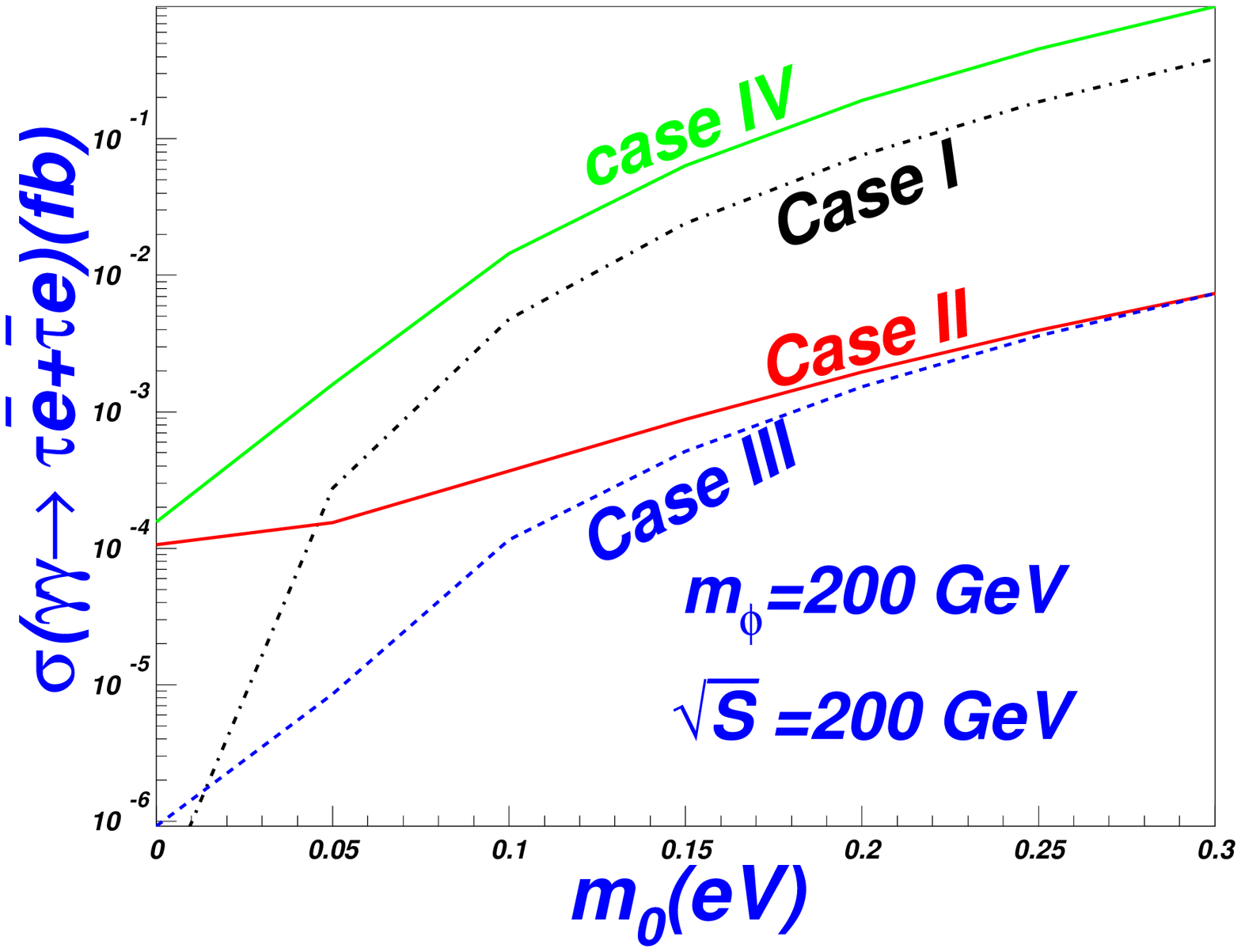,width=7.25cm} \figsubcap{b} }
 \parbox{7.05cm}{\epsfig{figure=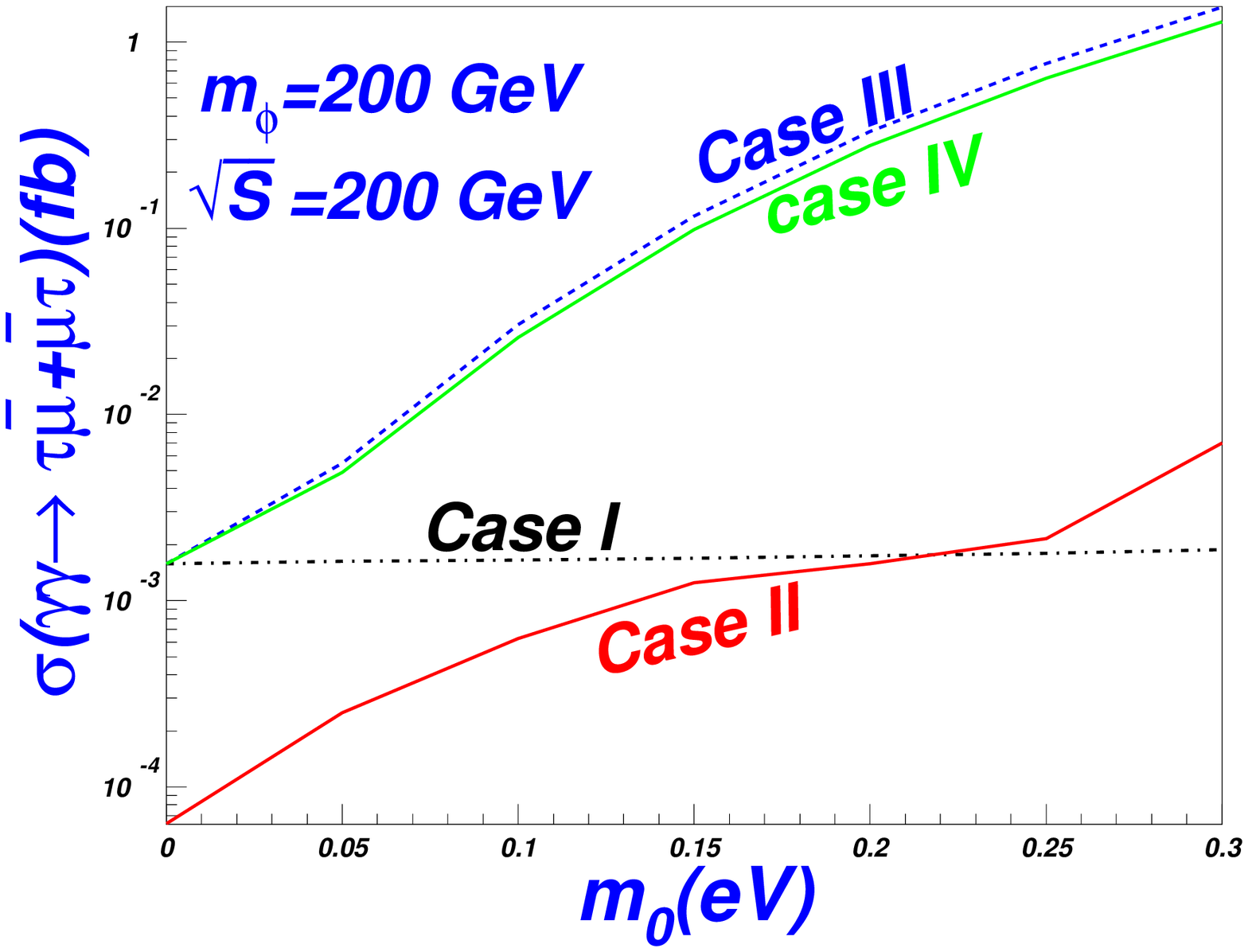,width=7.25cm} \figsubcap{c} }
 \caption{ The cross section $\sigma$ of the LFV process $\gamma\gamma
            \to   e\mu, ~\tau \bar e $ and $\tau\bar\mu$ as a function of the
 neutrino mass $m_0$ from  Case I to Case IV, 
  for the scalar mass $m_\phi=200$ GeV and $\sqrt{s}=200$ GeV,
 with the constraints $ h_{e\mu}= h_{ee}\approx 0$.
\label{fig9} }
\end{center}
\end{figure}

\begin{figure}[bht]%
\begin{center}
\hspace{-1.5cm}
 \parbox{8.25cm}{\epsfig{figure=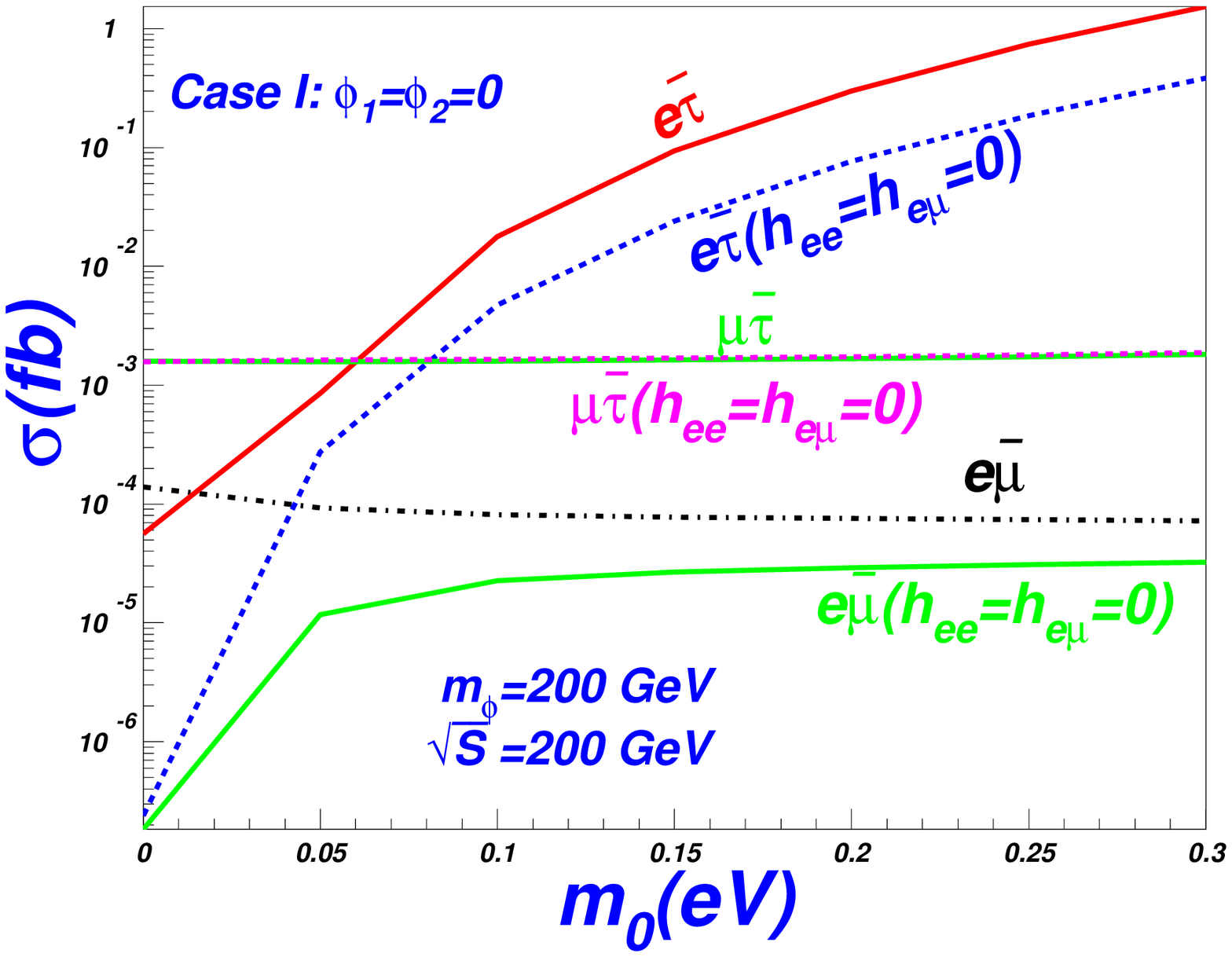,width=8.55cm} } 
 \parbox{8.25cm}{\epsfig{figure=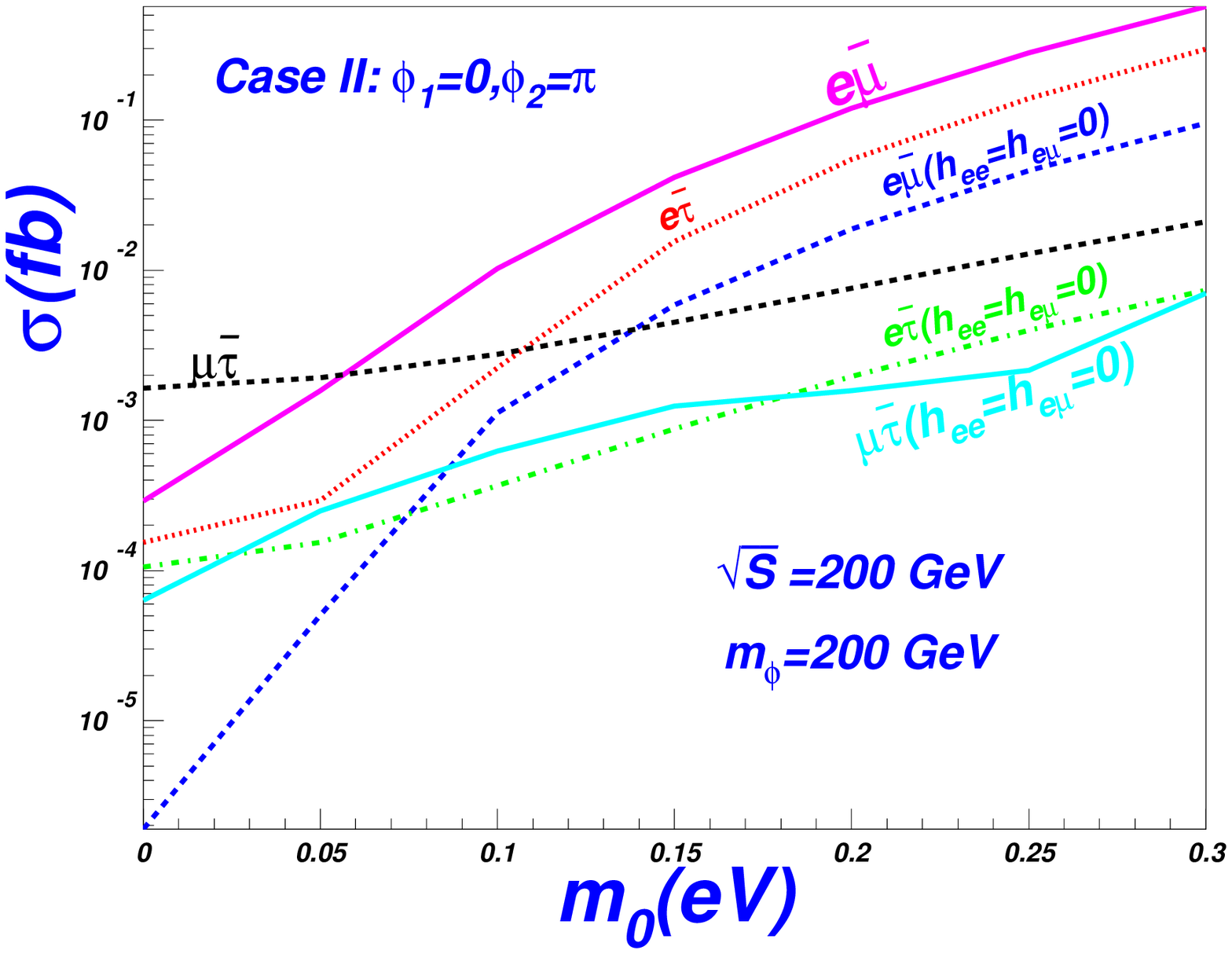,width=8.55cm} } 
 \end{center}
\begin{center}
\hspace{-1.5cm}
 \parbox{8.25cm}{\epsfig{figure=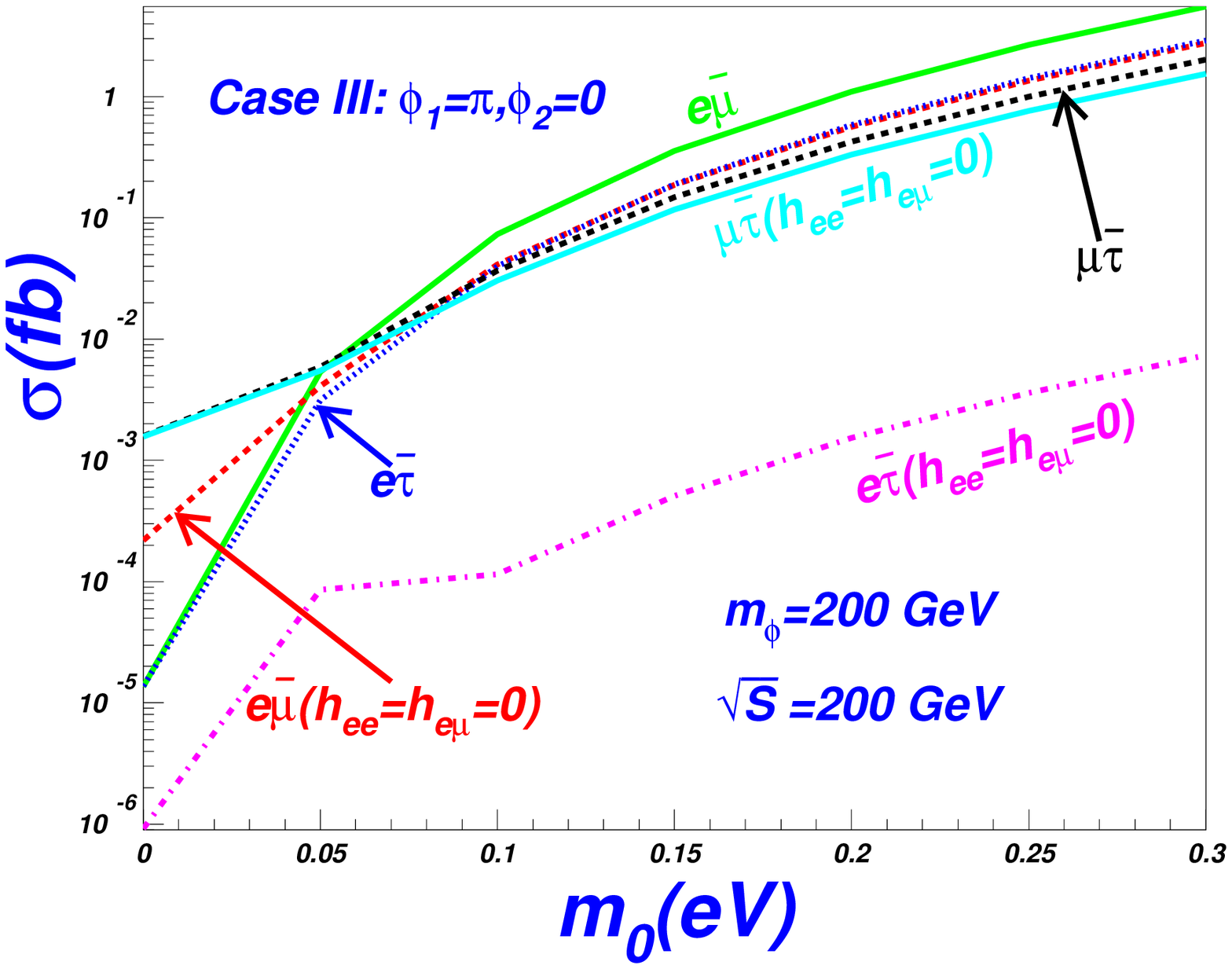,width=8.55cm} }
\parbox{8.25cm}{\epsfig{figure=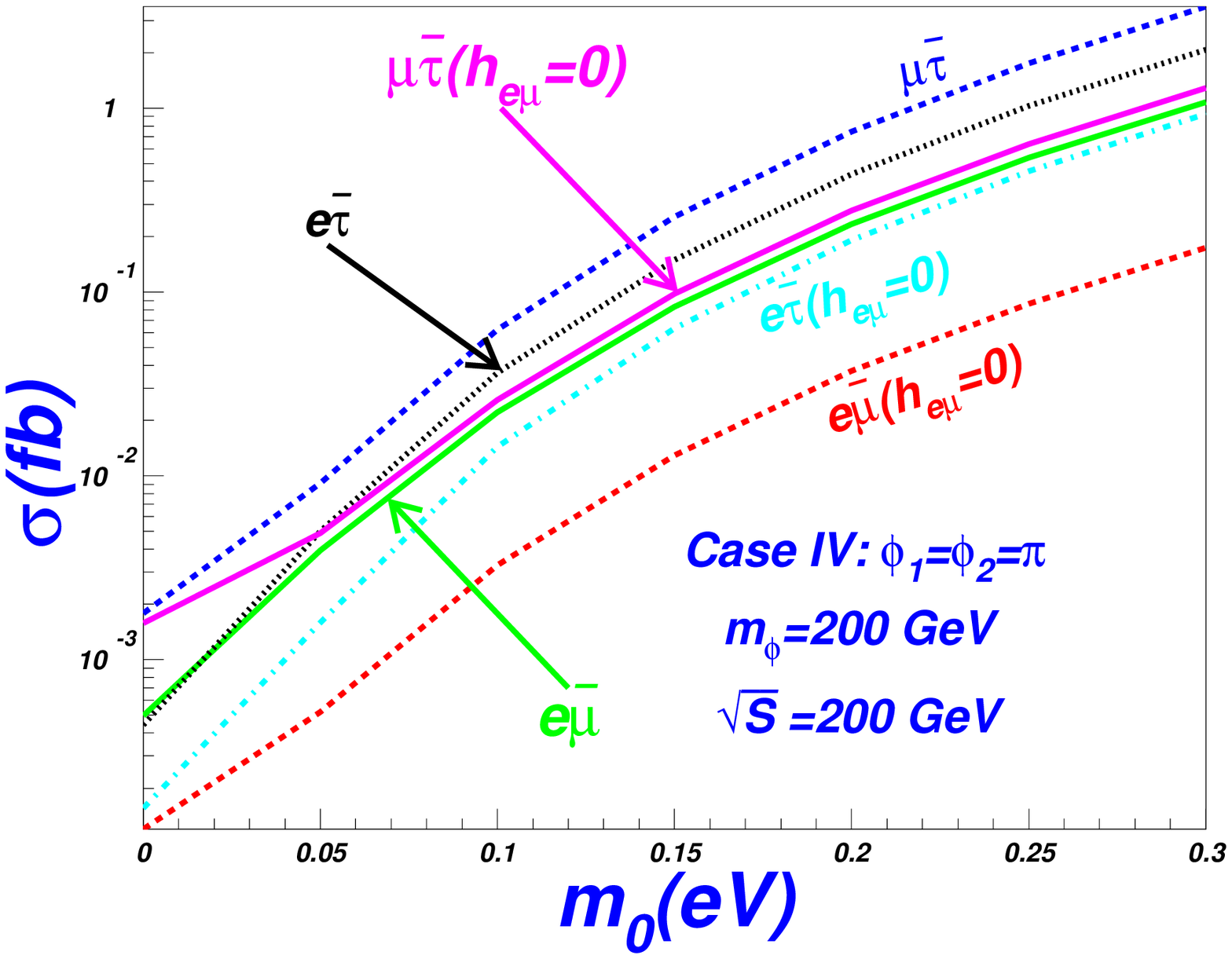,width=8.55cm} } 
  \caption{ Same as Figure \ref{fig9}, but compare the results with and without the constraints  $ h_{e\mu}= h_{ee}\approx 0$ in
  the same plots.
\label{fig10} }
\end{center}
\end{figure}

Besides, to see the results clearer, we also show the cross sections
with and without LFV constraints $ h_{e\mu}= h_{ee}\approx 0$ in Figure \ref{fig10}, from which, we can see that the production rates are
suppressed by the constraints.

\subsection{Scan $\phi_1$ and $\phi_2$ from $-\pi$ to $\pi$}
Since the $\phi_1$ and $\phi_2$ can affect greatly the relevant results, which can be seen from the four cases we have
discussed in Figures \ref{fig2}, \ref{fig3}, \ref{fig4}, \ref{fig5} and \ref{fig9}, we will scan the $\phi_1$ and $\phi_2$ from $-\pi$ to $\pi$ and see
the allowed range of $\phi_1$ and $\phi_2$ with the constraints $ h_{e\mu}= h_{ee}\approx 0$. This can be studied in both the normal hierarchy and inverted hierarchy.
We summarize the results in Table \ref{table1}. 

From Table \ref{table1}, we can see that the smallest cross sections for the LFV processes are quite small, especially when $m_0$ is smaller than $0.1$ eV and the values are less than $10^{-3}$ fb. From the scan, we can see that the smallest result for $\gamma\gamma\to \mu\bar e$ in the normal hierarchy is $0.00098$ fb, but most of the cross sections are larger than $0.1$ fb and can even reach $1$ fb in certain regions of the parameter space. So in much of the allowed parameter space, the cross sections are large. Besides, the allowed parameter space is not small.



\begin{table}[htb]
\begin{center}
\begin{tabular}{|c|c|c|c|c|c|c|} \hline
 $m_0$ & \multicolumn{3}{c|}{Normal Hierarchy}& \multicolumn{3}{c|} {Inverted Hierarchy} \\ 
\cline{2-7}
     (eV)   & $\mu \bar e $ & $\tau \bar e $ & $\tau \bar \mu $ &$\mu \bar e $          & $\tau \bar e $ & $\tau \bar \mu $  \\ \hline
  $0.25$ &(0.001,6.30)       &(0.0008,1.53)  & (0.002,3.66)  &(0.000013,1.38)    &(0.0003,0.44) &(0.002,1.83)  \\\hline
   $0.15$ &(0,0.186)  &(0.000001,0.06)& (0.00006,0.53)   & (0.000004,0.20) &(0.000003,0.04)&(0.002,0.53)\\\hline
$0.05$ &(0,0.004)  &(0,0.002)   & (0.00002,0.014)  &(0.000001,0.0056)  &(0.000002,0.002) &(0.00004,0.014)\\\hline
$0.00$ & (0,0.0002) & (0,0.000) &  (0,0.0017)     &(0,0.0003)& (0,0.0002)&(0.00004,0.002)\\\hline
\end{tabular}
\caption{Varying $-\pi < \phi_1,\phi_2 < \pi$,  approximate allowed ranges of the
cross sections of $\gamma\gamma \to l_i l_j$ ($i,j =e,~\mu,~\tau$) for several values of $m_0$ are shown.
Other parameters are the same as in Figure  \ref{fig2}.  We take the normal and the inverted neutrino mass hierarchy and the energy of ILC at $200$ GeV. The cross section are in the unit of fb and those too small are labeled as $0$.}
\label{table1}
\end{center}
\end{table}



\subsection{ the SM Backgrounds of the  $\gamma \gamma \to \ell_i \bar\ell_j$ }
With the following kinematical cuts \cite{susy-r-con1}: $|\cos\theta_e|<0.9$ and
$p^e_{T}>20{\rm ~GeV}$, the main SM backgrounds for
the process $\gamma \gamma \to \tau \bar e$ are
$\gamma\gamma \to \tau ^{+}\tau^{-} \to \tau \nu_{e}\bar{\nu}_{\tau}\bar e$,
$\gamma\gamma \to W^{+}W^{-}\to \tau \nu_\tau \nu_e \bar e $
and  $\gamma\gamma \to \tau \bar e \nu_\tau\nu_e$ which
 are suppressed to be $9.7\times 10^{-4}$ fb, $1.0\times 10^{-1}$ fb and
$2.4\times 10^{-2}$ fb \cite{susy-r-con1}, respectively.
Given $3.45 \times 10^2$ fb$^{-1}$ integrated luminosity of the photon collision\cite{tesla},
the production rates of $\gamma \gamma \to \mu\bar e,~\tau\bar{e}, ~\tau \bar{\mu}$
must be larger than $2.5\times 10^{-2}$ fb to get
the $3 \sigma$ observing significance \cite{susy-r-con1,rrmutau-susy}.

 We see from above Figures   \ref{fig4} and \ref{fig7}
  that under the current bounds that $h_{e\mu}\sim h_{ee} \sim 0$
  \cite{09094943,09043640,9511297,0304254,0304069} when the lightest neutrino mass is not too small, e.g, $m_0\geq 0.1$ eV,
the LFV process $\gamma\gamma\to \tau \bar e$ is large enough to enhance
 the production rate
 to $3 \sigma$ sensitivity and may be probed in the future ILC collider.

Unlike the process $ \ell_i \bar\ell_j$ production in supersymmetry\cite{susy-r-con1,rrmutau-susy}, the littlest Higgs with T-Parity\cite{11013598} and TC2 models\cite{10022607}, the cross section of the
$\gamma \gamma \to \mu \bar e $ is not definitely smaller than those of $\gamma \gamma \to \tau \bar \mu,~\tau\bar e $,
though the constraints from $\mu\to e\gamma$ and $\mu\to 3e$ give a quite small $h_{ee}$, $h_{e\mu}$. Because in the $e-u$ transition, even if we restrict the $h_{ee}$, $h_{e\mu}$ coupling to $0$, the other unsuppressed couplings $h_{\tau e},~h_{\tau \mu},~h_{\tau\tau}$ and  $h_{\mu\mu}$, induced by  all the three generations of the leptons which can enter the loop and contribute, could be large. Just as shown in Figure~\ref{fig1}, the leptons $\ell_k$ or $\nu_k$ ($k=1,2,3$) in the loop can make the couplings arbitrary unless some constraints are put on them.

For $\tau\bar\mu$ and $\mu \bar e$ production in the photon photon collision,
from Figures  \ref{fig2},  \ref{fig3},  and \ref{fig6}, we can see
that their cross sections are almost the same and a bit larger than
that of the $\tau e$ production. The SM backgrounds of them are the same if we
neglect the mass difference of the final lepton masses. So we can conclude
that detection of the $\tau\mu$ and $\mu e$ production may be advantageous than that of the $\tau e$.

\subsection{Comparison of the predictions of different models}
In this section, we first briefly recapitulate the sources of lepton flavor
violating transitions in different models and then compare the typical
magnitudes of various LFV  processes in the $\gamma\gamma$ collision predicted by different models.

It is well known that in the SM the LFV transitions are absent at
tree-level by the lepton number conservation.
The source of such LFV transitions  in the extensions of the
SM is the non-diagonality of the MNS  matrix.  These non-diagonal
elements can be large and may induce visible processes.

As the simplest extension of the SM, the HTM may naturally have LFV
mediated by the Higgs bosons at tree-level. In a popular realization
of HTM, the Higgs doublet is responsible for the electroweak
symmetry breaking as well as generating the fermion masses while
the triplet has LFV couplings whose strength are usually
parameterized by $h_{ij}$ which is shown in Eq. (\ref{hij_expressions}).

In the R-parity conservation MSSM 
 \cite{susy-r-conversation}, the neutrino masses can be obtained by introducing right-handed neutrinos and the non-diagonal
elements of the mass matrix give the LFV transitions like
$\ell_i \to \ell_j$, $i \neq j$, $\ell =e, ~\mu, ~\tau$ which are induced by
lepton-slepton-gaugino vertex through the diagnoliazation matrix $U_{Lij}$ \cite{susy-r-conversation} (or sneutrino mixing).

In the R-parity violating MSSM models \cite{r-violating-susy}, the lepton flavor changing (LFC)
couplings are provided by the $L$-violating coupling, with $L$ the lepton number, and the
bounds of the LFC couplings $\lambda$ and $\lambda'$ are given in TABLE I of the Ref. \cite{rrmutau-susy},
from which we can see the $\lambda$ and $\lambda'$ are constrained to be less than $10^{-2}$.

In the littlest Higgs model with T-parity (LHT) \cite{lht-lfv}, the interaction between the mirror lepton and the SM lepton,
such as $\bar l_H l Z_H(A_H)$ and  $\bar \nu_H l W_H$, can induce LFV interactions at loop level, that is, the new T-odd gauge bosons $Z_H, ~A_H, ~W_H$ can realize the transformation between different lepton in the loop level.

One of the dynamic EWSB models, the topcolor-assisted technicolor (TC2) model\cite{tc2-rev}, is quite different from the other models.
To tilt the chiral condensation in the $t\bar{t}$ direction and forbid the
formation of a $b\bar{b}$ condensation, a
non-universal extended $U(1)$ gauge group is needed in all TC2 models. Therefore, the existence
of the extra $U(1)$ gauge bosons $Z^{\prime}$ is predicted and such
new particle treats the third generation quarks and leptons differently from those in the first and second generations.
That is, it couples preferentially to the third generation fermions.
After the mass diagonalization from the flavor eigenbasis into the
mass eigenbasis, such new particle can lead to tree-level quark and lepton flavor changing couplings.

We conclude from Table \ref{compare} that, the
$\gamma \gamma$ collision is the better channel in enhancing the
magnitude for the $\ell_i\ell_j$ ($i\neq j$) associated productions at the ILC,
and the models listed there give sizable cross sections.
We can also see that, though the TC2 models generally predicts much larger LFV transitions than any other models,
 all these models can give large contributions and may be probed at the ILC.
 So even if we find some signal of the LFV processes, we also need to distinguish between these various new physics models. 

 As discussed in the former sections, motivated by the fact that
any process that is forbidden or strongly suppressed in the SM
constitutes a natural laboratory to search for new physics
effects, the LFV processes are of particular interests for us. It turns out that they may have large
cross sections, much larger than the SM ones, for certain models such as the MSSM, TC2 models and the HTM models.
We can see from Table \ref{compare} that the HTM model predicts LFV transition rates comparable to other new physics models predictions.


\begin{table}
\begin{center}
\begin{tabular}{|l|l|l|l|l|l|}
\hline
 &R-conversation MSSM &R-violating MSSM&TC2 &LHT &HTM \\ \hline
$\sigma(\gamma\gamma \to \tau\bar\mu)$    & ${\cal O}(10^{-2})$~\cite{susy-r-con1}  &${\cal O}(10^{-2})$ \cite{rrmutau-susy}  & ${\cal O}(1)$ \cite{10022607}     & ${\cal O}(1)$ \cite{11013598}& ${\cal O}(10^{-1})$ fb\\ \hline
$\sigma(\gamma\gamma \to\tau\bar e)$      & ${\cal O}(10^{-1})$~\cite{susy-r-con1} &${\cal O}(<10^{-1})$ \cite{rrmutau-susy} & ${\cal O}(1)$\cite{10022607}       & ${\cal O}(10^{-1})$\cite{11013598} & ${\cal O}(10^{-1})$ fb\\ \hline
$\sigma(\gamma\gamma \to \mu\bar e)$      & ${\cal O}(10^{-3})$~\cite{susy-r-con1} &${\cal O}(<10^{-3})$ \cite{rrmutau-susy} & ${\cal O}(10^{-3})$\cite{10022607} & ${\cal O}(10^{-1})$ \cite{11013598} & ${\cal O}(10^{-1})$ fb\\ \hline
\end{tabular}
\caption{Theoretical predictions for the $\ell_i\bar\ell_j$ ($i\neq j$) productions at $\gamma\gamma$
collision at the ILC. The predictions beyond SM are the optimum
values. The collider energy is $500$ GeV.}
\label{compare}
\end{center}
\end{table}
\section{Conclusion}
We have performed an analysis for the scalar-induced LFV productions
of $\ell_i\ell_j$ ($i\neq j$)  via $\gamma \gamma$ collision at
the ILC. We find that in the optimum part of the parameter space,
the production rate of $\gamma \gamma \to \ell_i\ell_j$ ($i\neq j$) can
reach $1$ fb. This means that we may have $100$ events each year for
the designed luminosity of $100$ fb$^{-1}$/year at the ILC. Since
the SM predictions of the production rates are completely negligible,
observation of such $\ell_i\ell_j$ events would be a
possible evidence of the HTM models. Therefore, these LFV processes
may serve as a sensitive probe of this kind of new physics models.
Since the LFV couplings are closely related to the neutrino masses, we may obtain interesting information for the neutrino masses from them if we could see
any signature of the LFV processes. At the same time, we compare the results of HTM with other new physics models and find that most predictions of these models can also be observed. So if we want to distinguish between these models through possible signals, further works are necessary.

\section*{Acknowledgments} \hspace{5mm}
This work was supported by the National Natural Science
Foundation of China under the Grants No.11105125, 11105124 and 11205023. 


\begin{thebibliography}{99}

\bibitem{Schechter:1980gr} J.~Schechter and J.~W.~F.~Valle,
Phys.\ Rev.\ D {\bf 22}, 2227 (1980). 

\bibitem{Cheng:1980qt}  T.~P.~Cheng and L.~F.~Li,
  Phys.\ Rev.\  D {\bf 22}, 2860 (1980).   

\bibitem{ilc-project} See the web: http://www.linearcollider.org/ILC/Publications/Reference-Design-Report.
\bibitem{clic-project} H. Braun et al., CLIC-NOTE-764, [CLIC Study Team Collaboration], CLIC 2008 parameters, http://www.clic-study.org.
\bibitem{rr-project} 
B. Badelek, et al, 	Int.J.Mod.Phys.A19 (2004), 5097-5186;
J. Gronberg, arXiv:1203.0031;
R. Nisius, arXiv:hep-ex/9811024;
A. Rosca, \EPJC33, (2004) s1044-s1046. 

\bibitem{09043640} 
A.G. Akeroyd, Mayumi Aoki, Hiroaki Sugiyama, \PRD79, (2009) 113010. 

\bibitem{Ma:2000wp}  E.~Ma, M.~Raidal and U.~Sarkar,
  Phys.\ Rev.\ Lett.\  {\bf 85}, 3769 (2000);
    E.~Ma, M.~Raidal and U.~Sarkar,
  Nucl.\ Phys.\ B {\bf 615}, 313 (2001).
\bibitem{0304069}
  E.~J.~Chun, K.~Y.~Lee and S.~C.~Park,
  Phys.\ Lett.\ B {\bf 566}, 142 (2003).
\bibitem{type2-roadmap}
A.~Melfo, M.~Nemevsek, F.~Nesti, G.~Senjanovic and Y.~Zhang,
\PRD85, (2012) 055018. 

\bibitem{9511297} 
J.A. Coarasa, A. Mendez, J. Sola, \PLB374, (1996) 131.



\bibitem{0304254} 
    M. Kakizaki, Y.  Ogura, F. Shima, \PLB566, (2003) 210.


\bibitem{Hahn} Hahn T, Perez-Victoria T,
     Comput.\ Phys.\ Commun., 1999, 118: 153-165;  
               Hahn T, 
     Nucl.\ Phys.\ Proc.\ Suppl.,\ 2004, 135: 333 


\bibitem{photon collider} Ginzburg I F, {\it et al.} 
Nucl. Instrum. Meth. A 1984, 219: 5-24.

\bibitem{pdg}  Amsler C, {\it et al.}, Particle Data Group.
 \PLB, 2008, 667: 1-5  and 2009 partial update for the 2010 edition.


\bibitem{0509152} 
C. A. de S. Pires, Mod.Phys.Lett. A21, (2006) 971. 

\bibitem{masiero}
 M. Lusignoli 1 A. Masiero, M. Roncadelli, Phys. Lett. B{\bf 252}, 247 (1990).

\bibitem{atlas} G. Aad et al. (ATLAS Collaboration), \PLB716, (2012) 1. 
\bibitem{cms} S. Chatrchyan et al.(CMS Collaboration), \PLB716, (2012) 30. 



\bibitem{07124019} 
A.G. Akeroyd, Mayumi Aoki, Hiroaki Sugiyama, \PRD77, (2008) 075010.
\bibitem{Maki:1962mu}
  Z.~Maki, M.~Nakagawa and S.~Sakata,
  Prog.\ Theor.\ Phys.\  {\bf 28}, 870 (1962).

\bibitem{Mphase}
  S.~M.~Bilenky, J.~Hosek and S.~T.~Petcov,
  Phys.\ Lett.\  B {\bf 94}, 495 (1980);
  M.~Doi, T.~Kotani, H.~Nishiura, K.~Okuda and E.~Takasugi,
  Phys.\ Lett.\  B {\bf 102}, 323 (1981).




\bibitem{Garayoa:2007fw}
  J.~Garayoa and T.~Schwetz,
  JHEP {\bf 0803}, 009 (2008). 


  \bibitem{Kadastik:2007yd}  M.~Kadastik, M.~Raidal and L.~Rebane,
  Phys.\ Rev.\  D {\bf 77}, 115023 (2008).

  \bibitem{Perez:2008ha}  P.~Fileviez Perez, T.~Han, G.~y.~Huang, T.~Li and K.~Wang,
  Phys.\ Rev.\  D {\bf 78}, 015018 (2008). 
\bibitem{solar}
  B.~T.~Cleveland {\it et al.},
  Astrophys.\ J.\  {\bf 496}, 505 (1998);
%
  W.~Hampel {\it et al.}  [GALLEX Collaboration],
  Phys.\ Lett.\  B {\bf 447}, 127 (1999);
%
  J.~N.~Abdurashitov {\it et al.}  [SAGE Collaboration],
  J.\ Exp.\ Theor.\ Phys.\  {\bf 95}, 181 (2002);
  [Zh.\ Eksp.\ Teor.\ Fiz.\  {\bf 122}, 211 (2002)]
%
  J.~Hosaka {\it et al.}  [Super-Kamkiokande Collaboration],
  Phys.\ Rev.\  D {\bf 73}, 112001 (2006);
%
  B.~Aharmim {\it et al.}  [SNO Collaboration],
  Phys.\ Rev.\ Lett.\  {\bf 101}, 111301 (2008);
%
  C.~Arpesella {\it et al.}  [The Borexino Collaboration],
  Phys.\ Rev.\ Lett.\  {\bf 101}, 091302 (2008)



\bibitem{atm}
  Y.~Ashie {\it et al.}  [Super-Kamiokande Collaboration],
  Phys.\ Rev.\  D {\bf 71}, 112005 (2005).
%
  J.L.~Raaf [Super-Kamiokande Collaboration],
a talk presented 23rd International Conference on Neutrino Physics
and Astrophysics (Neutrino 2008), Christchurch, New Zealand,
26-31 May 2008.


\bibitem{acc}
  M.~H.~Ahn {\it et al.}  [K2K Collaboration],
  Phys.\ Rev.\  D {\bf 74}, 072003 (2006);
%
  P.~Adamson {\it et al.}  [MINOS Collaboration],
  Phys.\ Rev.\ Lett.\  {\bf 101}, 131802 (2008)

  \bibitem{daya-bay} 
Daya Bay Collaboration, \PRL108, (2012) 171803; 
Chin.Phys. C37 (2013) 011001. 

\bibitem{phase-0nbb}  S.~Pascoli, S.~T.~Petcov and L.~Wolfenstein,
  Phys.\ Lett.\  B {\bf 524}, 319 (2002); 
  V.~Barger, S.~L.~Glashow, P.~Langacker and D.~Marfatia,
  Phys.\ Lett.\  B {\bf 540}, 247 (2002);
    H.~Nunokawa, W.~J.~C.~Teves and R.~Zukanovich Funchal,
  Phys.\ Rev.\  D {\bf 66}, 093010 (2002). 

\bibitem{09115291}S. A. Thomas, F. Abdalla, O. Lahav,
\PRL105, (2010) 031301. 

\bibitem{09094943} 
Takeshi Fukuyama (Ritsumeikan U., Kusatsu), Hiroaki Sugiyama (Ritsumeikan U., Kusatsu), Koji Tsumura (ICTP, Trieste)
\JHEP 1003, (2010) 044. 














\bibitem{susy-r-con1}
M. Cannoni, C. Carimalo, W. Da Silva, O. Panella, \PRD72, (2005) 115004; Erratum-ibid. D72, (2005) 119907. 
\bibitem{tesla} Badelek B {\it et al.}, 
    Int.\ J.\ Mod.\ Phys.\  A, 2004, 19: 5097-5186. 

\bibitem{rrmutau-susy} Cao J, Wu L, Yang J. 
\NPB, 2010, 829: 370-382;  
Sun Yan-Bin, Han Liang, Ma Wen-Gan, Tabbakh Farshid, Zhang Ren-You, Zhou Ya-Jin,  \JHEP0409, (2004) 043,2004. 


\bibitem{11013598}
Jin-zhong Han, Xue-lei Wang, Bing-fang Yang, \NPB843,(2011) 383.

\bibitem{10022607}
Guo-Li Liu,  Science China, 53(2010)1-6. 

\bibitem{susy-r-conversation}F. Borzumati and A. Masiero, Phys. Rev. Lett. 57, (1986) 961; J. Hisano, T. Moroi, K. Tobe, M. Yamaguchi, Phys. Rev. D 53, (1996) 2442; J. Hisano and D. Nomura, Phys. Rev. D 59, (1999) 116005.




\bibitem{r-violating-susy} For some early works on R-violating supersymmetry, see, e.g.,
C.~S.~Aulakh and R.~N.~Mohapatra,  
 \PLB119, (1982) 136;
L. Hall and M. Suzuki, \NPB231, (1984)419; J. Ellis et al., Phys. Lett. B 150, 142 (1985); G. Ross and J. Valle,
\PLB151, 375 (1985); S. Dawson, \NPB 261, 297 (1985); R. Barbieri and A.
Masiero, \NPB267, 679 (1986); H. Dreiner and G.G. Ross, \NPB 365, 597
(1991); J. Butterworth and H. Dreiner, \NPB 397, 3 (1993).


\bibitem{lht-lfv} I. Low, \JHEP, 0410, 067(2004); H. C. Cheng and I. Low, \JHEP, 0408, (2004) 061; J. Hubisz and P. Meade, \PRD71, (2005) 035016; J. Hubisz, S. J. Lee and G. Paz, \JHEP, 0606, (2006) 041;
M. Blanke, A. J. Buras, A. Poschenrieder, Recksiegel C. Tarantino, S. Uhlig and A. Weiler, \JHEP0611, (2006) 062.

\bibitem{tc2-rev}Hill C T. 
 \PLB, 1995, 345: 483-489;  
            Lane K and Eichten E. 
 \PLB, 1995, 352: 382-387;  
            Lane K. 
\PLB, 1998, 433: 96-101;  
            Cvetic G. 
\RMP, 1999, 71: 513 

S. Wang et al., \PRD74, (2006) 057902.


\end{thebibliography}
\end{document}